\documentclass{emulateapj}

\usepackage{color}

\shorttitle{}
\shortauthors{}

\begin{document}

\title{Criteria for Core-Collapse Supernova Explosions by the Neutrino Mechanism}

\author{Jeremiah W. Murphy\altaffilmark{1,2,3}}
\author{Adam Burrows\altaffilmark{4,1}}

\altaffiltext{1}{Steward Observatory, The University of Arizona,
  Tucson, AZ 85721; jmurphy@as.arizona.edu}
\altaffiltext{2}{Astronomy Department, University of Washington, Box
  351580, Seattle, WA 98195-1580; jmurphy@astro.washington.edu}
\altaffiltext{3}{NSF Astronomy and Astrophysics Postdoctoral Fellow}
\altaffiltext{4}{Department of Astrophysical Sciences, Princeton University,
  Princeton, NJ 08544; burrows@astro.princeton.edu}

\begin{abstract}
We investigate the criteria for successful
core-collapse supernova explosions by the neutrino mechanism.
We find that a critical-luminosity/mass-accretion-rate
condition distinguishes non-exploding from exploding models in
hydrodynamic one-dimensional (1D) and two-dimensional (2D)
simulations.  We present 95 such simulations that parametrically explore
the dependence on neutrino luminosity, mass accretion rate, resolution, and dimensionality.
While radial oscillations mediate the transition between 1D accretion
(non-exploding) and exploding simulations, the non-radial standing
accretion shock instability characterizes 2D simulations. We find 
that it is useful to compare the average dwell time of matter in the
gain region with the corresponding heating timescale, but that tracking the 
residence time distribution function of tracer particles better describes
the complex flows in multi-dimensional simulations.  Integral
quantities such as the net heating rate, heating efficiency, and mass
in the gain region decrease with time in non-exploding models, but for
2D exploding models, increase before, during, and after explosion.
At the onset of explosion in 2D, the heating efficiency is $\sim$2\% to
$\sim$5\% and the mass in the gain region is $\sim$0.005
M$_{\sun}$ to $\sim$0.01 M$_{\sun}$.  Importantly, we find that
the critical luminosity for explosions in 2D is $\sim$70\%
of the critical luminosity required in 1D.
This result is not sensitive to resolution or whether the 2D
computational domain is a quadrant or the full 180$^{\circ}$.
We suggest that the relaxation of the explosion condition in going
from 1D to 2D (and to, perhaps, 3D) is of a general character and is 
not limited by the parametric nature of this study.
\end{abstract}


\section{Introduction}

Four decades of core-collapse simulations have increased our
understanding of the core-collapse mechanism, yet a complete
theory of the mechanism has not emerged.  Detection of neutrinos during SN 1987A
\citep{bionta87,hirata87} confirmed only the fundamentals of
core-collapse supernovae.  Theory suggests that the Fe core collapses to form a protoneutron
star (PNS), which launches a shock wave.  Before the bounce
shock can explode the star, it is
sapped of energy by nuclear dissociation and neutrino losses and
stalls into an accretion shock \citep{mazurek82,bruenn85,bruenn89}.
Understanding the mechanism that revives the stalled shock into
explosion has been the goal of the community for many decades.

Since the pioneering work of
\citet{wilson85} and \citet{bethe85}, the favored shock revival
mechanism has been the delayed-neutrino mechanism (or simply neutrino mechanism), in which neutrinos
heat the post-shock region and restart the shock's outward progress
after hundreds of milliseconds of delay.  Detailed one-dimensional (1D) simulations using state-of-the-art equations of state
(EOSs), neutrino-matter cross sections, and neutrino transport have
shown that the neutrino mechanism fails to produce explosions in 1D
\citep{liebendorfer01a,liebendorfer01b,rampp02,buras03,thompson03,liebendorfer05b},
except the least massive of the massive stars \citep{kitaura06,burrows07d}. 
Recent 2-dimensional (2D) simulations, and the accompanying
aspherical instabilities, have suggested that the neutrino mechanism may
yet succeed, though it fails in 1D
\citep{herant94,janka95,burrows95,janka96,burrows07a,kitaura06,buras06a,buras06b,marek07,ott08}.
Thus, the fundamental question of core-collapse theory remains; how is
accretion reversed into explosion?

Exposing the core-collapse mechanism will
require detailed three-dimensional (3D) radiation-hydrodynamic
simulations.  However, the core-collapse problem is messy,
with many subtle nonlinear couplings and feedbacks 
on local and global scales, and extracting the essence of the mechanism from even 1D radiation-hydrodynamic simulations has
proven very difficult.  Hence, revealing the core ingredients and conditions of the mechanism will
likely require a
two-front attack.  On the one hand, multi-dimensional
radiation-hydrodynamic simulations will give fully consistent
explosions, producing observationally testable energies,
neutron star masses, nucleosynthetic yields, and more.
On the other hand, simplified models that nevertheless retain the important physics,
but allow adjustment of important parameters will help reveal what is important.  In this paper, we pursue the latter
philosophy.  We present a systematic parameterization of the conditions
and criteria for explosion by the neutrino mechanism, emphasizing the
effect that going to 2D from 1D has on the critical neutrino luminosity
required for explosions.

\citet{burrows93} suggested a simple framework for determining the
conditions for successful explosions by the neutrino mechanism.
They approximated the stalled shock and accretion phase as a steady-state problem, transforming the governing partial differential
equations into ordinary differential equations.  By
parameterizing the electron-neutrino luminosity, $L_{\nu_e}$, and the
mass accretion rate, $\dot{M}$, they identified a critical $L_{\nu_e}$-$\dot{M}$ curve that
distinguishes steady
state accretion solutions (lower luminosities) from explosions (high luminosities).  This implied
that global conditions, not local conditions, mediate the transition
from accretion to explosion, which in turn suggests that core-collapse
explosion is a global instability.
More than a decade later, \citet{yamasaki05,yamasaki06}
reproduced these results and performed a
linear stability analysis of the steady-state solutions that showed
unstable solutions near the critical luminosity.
In this paper, 
we extend this previous work by performing a suite of hydrodynamic
simulations in both 1D and 2D.

Assuming that the concept of a critical luminosity applies to 2D simulations, many have noted that critical luminosities for 2D
simulations are likely to be lower than in 1D simulations.  In their 1D steady-state
solutions, \citet{yamasaki05,yamasaki06} investigated the
effects of rotation and convection on the critical luminosity, but
these analyses lacked a self-consistent treatment of what is an inherently 2D/3D phenomenon.
Early 2D simulations using flux-limited diffusion noted a trend toward
explosions aided by neutrino-driven convection
\citep{burrows95,janka96}, while 1D simulations failed to
explode for the same neutrino luminosity.  More recently,
\citet{blondin03} identified a new instability that may prove crucial
for the viability of the neutrino mechanism.  It is the standing accretion
shock instability (SASI), which may be an advective-acoustic \citep{foglizzo00,foglizzo02,foglizzo07} or purely
acoustic cycle \citep{blondin06}.  Whichever explains the SASI, recent simulations have suggested
that it may facilitate the neutrino
mechanism \citep{marek07,buras06b}.  However, these simulations have yet
to demonstrate a reliable explosion mechanism for a wide range of progenitor
masses that yields explosion energies consistent with Nature.
Furthermore \citet{burrows06}, \citet{burrows07a},
\citet{burrows07b}, \citet{burrows07d}, and \citet{burrows07c}, using
VULCAN/2D \citep{livne93,livne04} and multi-group
flux-limited diffusion (MGFLD), obtain
successful explosions only for the 8.8-M$_{\sun}$ model by the
neutrino mechanism alone.
Recently, \citet{ott08} compared MGFLD and multi-angle transport in VULCAN/2D.  While they note some interesting
differences between the transport schemes, explosions remain elusive.
Moreover, it
has not been demonstrated that the concept of a critical
$L_{\nu_e}$-$\dot{M}$ condition pertains to 2D simulations, and if it
does, how do the critical luminosities of 1D and 2D simulations compare?

To address these questions, we conduct 1D and 2D simulations for
various values of $L_{\nu_e}$.  In the past, there have been a few
investigations on the systematic effects of neutrino luminosity on the
explosion mechanism \citep{janka96}, neutron star kicks
\citep{scheck06}, and the SASI \citep{ohnishi06}, but none has
thoroughly investigated both $L_{\nu_e}$ and $\dot{M}$ in 1D and 2D
simulations to address the viability of a critical luminosity condition
for explosion.
In addition to $L_{\nu_e}$ and $\dot{M}$, we compare simulations with
different spatial
resolutions.  Using the code, BETHE-hydro \citep{murphy08},
we simulate the core-collapse, bounce, and post-bounce phases in
time-dependent 1D and 2D simulations.  These simulations have no
inner boundary and include the PNS core.  A
finite-temperature EOS
that accounts for nucleons, nuclei,
photons, electrons, positrons, and all the appropriate phase
transitions is used \citep{shen98}.  Employing 11.2- and 15-M$_{\sun}$ progenitors \citep{woosley02,woosley95} as initial conditions, a wide range of $\dot{M}$ is
sampled (from $\sim$0.08 M$_{\sun}$/s to $\sim$0.3 M$_{\sun}$/s).
Finally, we use standard
approximations for neutrino heating and cooling that enable a
straightforward parameterization of $L_{\nu_e}$ \citep{bethe85,janka01}.

The basic equations and the numerical techniques are presented in \S
\ref{section:numerics}.
In \S \ref{section:progenitor}, we discuss the progenitor models
and describe the suite of simulations performed.  In \S
\ref{section:shock}, we analyze the evolution of the shock for 1D and 2D
simulations and discuss the role shock oscillations play in the explosion.  In \S \ref{section:grid}, we
investigate the effect that the grid, specifically spatial resolution and
the angular extent of the domain, has on explosions.  In \S
\ref{section:timescales}, we revisit the
condition for explosion as expressed by the heating and advection
timescales, and in \S \ref{section:conditions} describe other
conditions at explosion.  In \S \ref{section:critical}, we quantify
the differences in the critical luminosity 
condition between 1D and 2D simulations.
Finally, in \S \ref{section:conclusions}, we summarize our conclusions.

\section{Equations and Numerical Techniques}
\label{section:numerics}

The basic equations of hydrodynamics are
the conservation of mass,
momentum, and energy: 
\begin{equation}
\label{eq:mass_lag}
\frac{d \rho}{d t} = - \rho \vec{\nabla} \cdot \vec{v} \, ,
\end{equation}
\begin{equation}
\label{eq:mom_lag}
\rho \frac{d \vec{v}}{d t} = - \rho \vec{\nabla} \Phi - \vec{\nabla} P
\, ,
\end{equation}
and
\begin{equation}
\label{eq:ene_lag}
\rho \frac{d \varepsilon}{d t} = - P \vec{\nabla} \cdot \vec{v} 
+ \mathcal{H} - \mathcal{C} \, .
\end{equation}
$\rho$ is the mass density, $\vec{v}$ is the fluid velocity, $\Phi$ is
the gravitational potential, $P$ is the isotropic pressure,
$\varepsilon$ is the specific internal energy, and $d/dt = \partial
/ \partial t + \vec{v} \cdot \vec{\nabla}$ is the Lagrangian time
derivative.
In this work, the neutrino heating, $\mathcal{H}$, and cooling, $\mathcal{C}$, terms
in eq. (\ref{eq:ene_lag}) are assumed to be
\begin{equation}
\label{eq:heating}
\mathcal{H} = 1.544 \times 10^{20} L_{\nu_e} \left ( \frac{100 {\rm
      km}}{r} \right )^2 \left ( \frac{T_{\nu_e}}{4 {\rm Mev}} \right
)^2 \left [ \frac{\rm erg}{{\rm g} \, {\rm s}} \right ] \, ,
\end{equation}
and
\begin{equation}
\label{eq:cooling}
\mathcal{C} = 1.399 \times 10^{20} \left ( \frac{T}{2 {\rm MeV}}
\right )^6
\left [ \frac{\rm erg}{{\rm g} \, {\rm s}} \right ] \, .
\end{equation}
Note that these approximations for heating and cooling by
neutrinos \citep{bethe85,janka01} depend upon local quantities and
predefined parameters.  They are $\rho$, temperature ($T$),
the distance from the center ($r$), the electron-neutrino temperature ($T_{\nu_e}$), and the electron-neutrino luminosity ($L_{\nu_e}$), which
is in units of $10^{52}$ erg s$^{-1}$.  By using
eqs. (\ref{eq:heating}) and (\ref{eq:cooling}), we gain considerable
time savings by approximating the effects of detailed neutrino transport.  For
all simulations, we set $T_{\nu_e} = 4$ MeV.  In eq. (\ref{eq:heating}), it has
been assumed that $L_{\nu_e} = L_{\bar{\nu}_e}$ and that the mass
fractions of protons and neutrons sum to one.  Therefore, the sum of
the electron- and anti-electron-neutrino luminosities is $L_{\nu_e
  \bar{\nu}_e} = 2 L_{\nu_e}$.
Closure for eqs. (\ref{eq:mass_lag}-\ref{eq:ene_lag}) is obtained
with an EOS appropriate for matter in or near nuclear statistical
equilibrium (NSE) \citep{shen98}, and the effects of photons, electrons,
and positrons are included. As such, the
EOS has the following dependencies:
\begin{equation}
\label{eq:eos}
P = P(\rho,\varepsilon,Y_e) \, ,
\end{equation}
where $Y_e$ is the electron fraction.
Therefore, we also solve the equation:
\begin{equation}
\label{eq:advec}
\frac{d Y_e}{dt} = \Gamma_e \, ,
\end{equation}
where $\Gamma_e$ is the net rate of $Y_e$ change.

Using BETHE-hydro
\citep{murphy08}, we solve eqs. (\ref{eq:mass_lag}-\ref{eq:ene_lag}) in one and two
dimensions by the Arbitrary Lagrangian-Eulerian (ALE) method. To
advance the discrete equations of hydrodynamics by one timestep, ALE
methods generally use two operations, a Lagrangian hydrodynamic step
followed by a remap.  The structure of BETHE-hydro's
hydrodynamic solver is designed for arbitrary-unstructured grids, and the
remapping component offers control of the time evolution of the grid.
Taken together, these features enable the use of
time-dependent arbitrary grids to avoid some unwanted features of
traditional grids.

For the calculations presented in this paper, we use this flexibility
to avoid the singularity of spherical grids in two dimensions.  While spherical
grids are generally useful for core-collapse simulations, the
convergence of grid lines near the center place extreme constraints on
the timestep via the Courant-Friedrichs-Levy condition.  A common remedy is to
simulate the inner $\sim$10 km in 1D or to use an inner boundary
condition.   Another approach, which has been used in VULCAN/2D simulations \citep{livne93,livne04}, is to avoid the singularity with a grid
that is pseudo-Cartesian near the center and smoothly transitions to a
spherical grid at larger radii.  We use a similar grid, the butterfly
mesh \citep{murphy08}, for the
simulations in this paper.

For 1D simulations, 700 radial zones are distributed from the center
to 4000 km.  The innermost 100
zones have a resolution of 0.34km, and the remaining 600 zones are
spaced logarithmically from 34 km to
4000 km.  After each Lagrangian hydrodynamic solve, the flow is
remapped back to the original grid, which in effect produces an
Eulerian calculation.  For 2D simulations, we use a butterfly mesh interior to 34 km and a
spherical grid exterior to this radius.  For all 180$^{\circ}$
simulations, the innermost pseudo-rectangular region is 50 by 100
zones, and the region that transitions from Cartesian to spherical
geometry has 50 radial zones
and 200 angular zones.  The outermost spherical region has 200 angular
zones and 300 or 150 radial zones which are logarithmically spaced
between 34 km and either 4000 km or 1000 km.  The butterfly portion has an
effective radial resolution of 0.34 km with the shortest cell edge
being 0.28 km.  The grids for the 90$^{\circ}$ simulations are similar,
but with half the number of zones.  In 2D simulations, we do not remap every
timestep.  Since the timestep is limited by the large sound speeds in
the PNS core, we can afford to perform several Lagrangian hydrodynamic
solves before remapping back to the original grid.  By remapping
seldomly, we gain considerable savings in computational time.

We investigate the systematics of the neutrino mechanism by parameterizing
$L_{\nu_e}$, $\dot{M}$, resolution, and the dimensions of the simulation.
Since this requires a large set of simulations, we make
  several approximations to expedite the calculations. 
For one, gravity is calculated via $\vec{g} = -G M_{\rm int}/r^2$, where $M_{\rm int}$ is
    the mass interior to the radius $r$. 
During collapse, rather than solving the rate equations for $Y_e$ due to
    neutrino transport, we use a $Y_e$ vs. $\rho$ parameterization.  \citet{liebendorfer05a}
observed that 1D simulations including neutrino transport produce
$Y_e$ values during collapse which are essentially a function of
density alone.  This allows for a parameterization of
$Y_e$ as a function of $\rho$.  To change $Y_e$, we use results of 1D SESAME
\citep{burrows00,thompson03} simulations to define the function
$Y_e(\rho)$, and we employ the
prescription of \citet{liebendorfer05a} to calculate local values of $\Gamma_e$.  As a result, the most
important effects of electron capture during collapse
are included without the need for expensive, detailed neutrino transport.

\section{Progenitor and Simulation Models}
\label{section:progenitor}

\begin{figure}
\plotone{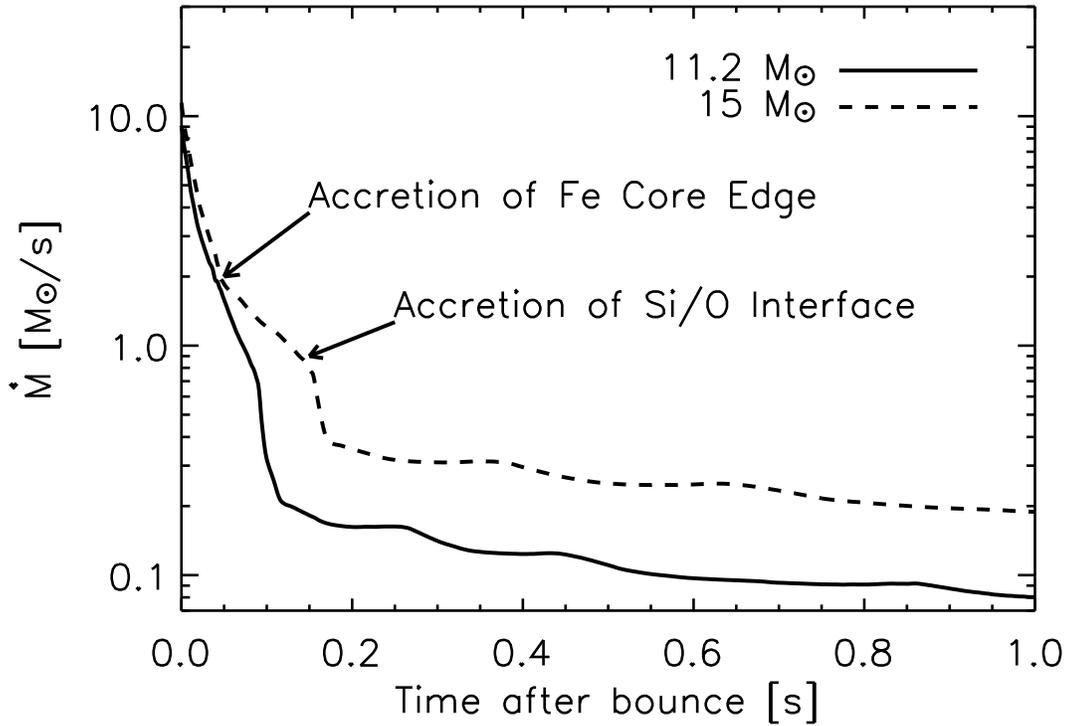}
\caption{Accretion rate, $\dot{M}$, vs. postbounce time above the
  stalled shock (250 km). The solid and dashed lines show the
  time-dependent accretion rate for the 11.2- and 15.0-M$_{\sun}$ models,
  respectively \citep{woosley02,woosley95}.  While the outer portions
  of the Fe core accrete ($t = 0$-50 ms), $\dot{M}$ is as high as 10
  M$_{\sun}$/s and decreases to 2 M$_{\sun}$/s.  After the Fe core fully
  accretes, the accretion rates for the two models
  diverge.  For the 11.2-M$_{\sun}$ (15-M$_{\sun}$) model, it takes 50 ms (100 ms) to accrete the
  Si-burning shell.
As the Si/O interface accretes, $\dot{M}$ plummets to 0.3 M$_{\sun}$/s
(0.2 M$_{\sun}$/s) for the 15 M$_{\sun}$ (11.2 M$_{\sun}$) model.  Afterward,
$\dot{M}$ declines slowly to 0.2 M$_{\sun}$/s (0.08 M$_{\sun}$/s).
Together, these two models slowly sweep through a
  range of accretion rates from 0.3 M$_{\sun}$/s to 0.08 M$_{\sun}$/s,
  which enables a parameterization in $\dot{M}$. \label{mdotvstime}}
\end{figure}

To sample a range of $\dot{M}$, we initiate the simulations with
two core-collapse progenitor models that have different density profiles: a
15 M$_{\sun}$ progenitor with a shallow density profile exterior to the Fe
core \citep{woosley95}, and a 11.2 M$_{\sun}$ progenitor with a steeper
exterior density profile \citep{woosley02}.
In Fig. \ref{mdotvstime}, the mass accretion rate, $\dot{M}$, vs. time at 250 km (just above the
  stalled shock) is shown for 1D non-exploding models. The solid and dashed lines show the
  time-dependent accretion rate for the 11.2 and 15.0 M$_{\sun}$ models.
  While the outer portions of the Fe core accrete through the shock, $\dot{M}$ is as high as 10
  M$_{\sun}$/s and decreases to 2 M$_{\sun}$/s within 50 ms of bounce for both
  masses.  After the Fe core is fully
  accreted, the accretion rates for the two models
  deviate.
It takes 50 ms to fully accrete the
  Si-burning shell using the 11.2-M$_{\sun}$ model and 100 ms using
  the 15-M$_{\sun}$ model.
Afterward,
$\dot{M}$ declines slowly to 0.2 M$_{\sun}$/s (0.08 M$_{\sun}$/s for the 15
M$_{\sun}$ model).
Together, these two models slowly sweep through a
  range of accretion rates from 0.3 M$_{\sun}$/s to 0.08 M$_{\sun}$/s,
  enabling a study of the $\dot{M}$ dependence.

The results presented in the following sections are derived from 95
simulations that as a group, represent a
parameterization of $L_{\nu_e}$, $\dot{M}$, resolution, and dimensionality.
Table \ref{table:sequences} lists these simulations into 14 sequences.  Each sequence of simulations is
distinguished by the progenitor model (column 2), dimensionality
(column 3), number of radial zones (column 4), radius of the outer
boundary (column 5), and a range of electron-neutrino luminosities
(column 6).  The sequence names encode the information contained in columns 2
through 5.  The first
four characters indicate the progenitor model.  Next are the
dimensionality labels:  ``1D'' for 1D simulations, ``2D'' for 2D
simulations that incorporate the full of $\theta$ (180$^{\circ}$),
and ``Q'' for 2D simulations that simulate the quadrant that lies
between the pole and equator (90$^{\circ}$).
For the 2D simulations, the final character represents the
resolution:  ``1'' for the lowest resolution of 250 effective radial
zones within 4000 km, ``2'' for a finer resolution of 400 effective
radial zones in 4000 km, and ``3'' for the highest resolution of 400
effective radial zones within 1000 km.  On the whole, 
Table \ref{table:sequences} represents a parameter study for the
conditions near explosion in $L_{\nu_e}$, $\dot{M}$, dimensionality, and
resolution.

\section{Radial Shock Oscillations in 1D and the SASI in 2D}
\label{section:shock}

When \citet{burrows93} reported a critical luminosity vs.
accretion rate condition for successful explosions, they did so based
upon steady-state solutions.
It was left to subsequent calculations to show that
the concept of a critical luminosity condition was
appropriate for time-dependent simulations.
With plots of the evolution of the shock radius, we show that a critical
luminosity condition distinguishes non-exploding and exploding models
in time-dependent 1D and 2D simulations, and that the
  critical luminosity for 2D simulations is lower than
  the critical luminosity for 1D simulations.
In addition, we report global oscillations near the
  critical luminosity in both 1D and 2D simulations, but we show that
  the oscillations are of a quite different character.

\begin{figure}
\epsscale{0.7}
\plotone{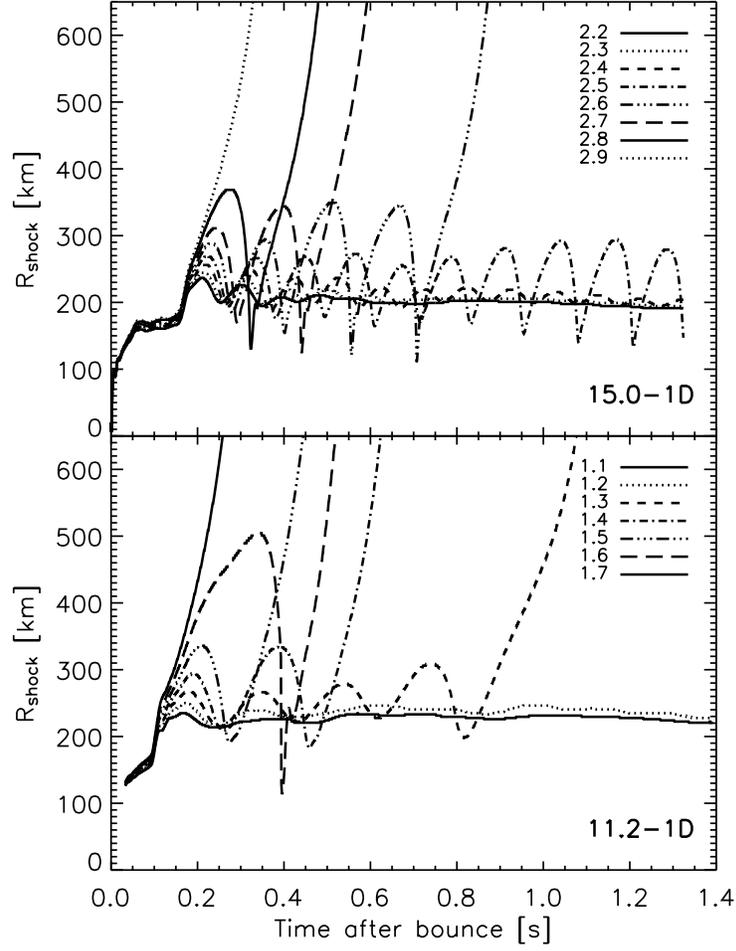}
\caption{Shock radius, $R_{\rm
  shock}$, vs. postbounce time for the 1D
sequences.  The top panel displays the radii for 15.0-1D, and the
bottom panel shows the radii for 11.2-1D.  Each line is labeled by the
electron-neutrino luminosity in units of $10^{52}$ ergs s$^{-1}$.
Comparing models within a sequence, all models show similar behavior prior to the accretion of
the Si/O interface.  The accretion of the Si/O interface either initiates an
explosion or excites radial oscillations in the shock radius.  For the
15.0-1D sequence, the oscillation periods range from $\sim$90 ms
for $L_{\nu_e}=2.2$ to $\sim$170 ms for $L_{\nu_e} = 2.8$.  Of the
models that oscillate, the lower luminosity simulations decay in
the oscillation amplitude, while higher luminosity models oscillate
and explode.  The timescales for decay range from $\sim$450 ms for
$L_{\nu_e} = 2.2$ to 1.0 s for $L_{\nu_e} = 2.5$.
\label{rvst_1d}}
\end{figure}

The time evolution for the shock radius, $R_{\rm shock}$, of 1D
simulations is presented in Fig. \ref{rvst_1d}.
In the top panel, we display the radii for many models of the 15.0-1D
sequence, and each is labeled by $L_{\nu_e}$.  The range of
luminosities shown, $L_{\nu_e} = 2.2$-2.9 (in units of $10^{52}$ erg s$^{-1}$), highlight values near the critical luminosity for
this model.  The fact that several luminosities lead to explosion at
different times is an early indication that the critical luminosity depends upon
the mass accretion rate. In \S \ref{section:critical}, we
elaborate on the accretion rate dependence.  Similarly, the bottom
panel shows $R_{\rm shock}$ for the range of luminosities surrounding
the critical luminosity, $L_{\nu_e}
= 1.1$-1.7, of the 11.2-1D sequence.

Prior to the accretion of
the Si/O interface, the shock stalls for both sequences at radii that are
slightly dependent upon $L_{\nu_e}$.
The accretion of this interface and the abrupt drop in $\dot{M}$
trigger an immediate explosion for the highest luminosity, $L_{\nu_e}
= 1.7$ (2.9) for the 11.2-1D (15.0-1D) sequence and
oscillations in $R_{\rm shock}$ for the other luminosities considered.
These radial oscillations either decay or maintain large amplitudes
until the model explodes.  As an example, the oscillation periods of
the 15.0-1D sequence range from $\sim$90 ms
($L_{\nu_e}= 2.2$) to $\sim$170 ms ($L_{\nu_e} = 2.8$), and the
timescales for decay range from 450 ms ($L_{\nu_e} = 2.2$) to 1.0 s
($L_{\nu_e} = 2.5$).  These radial oscillations and their large
amplitudes near explosion support, but do not prove, the hypothesis
that a global instability is responsible for the transition between
accretion and explosion.

Near the critical luminosity, others have reported large radial oscillations in 1D simulations.
\citet{ohnishi06} executed time-dependent 1D simulations of the
    steady state solutions in \citet{yamasaki05} and observed similar
    pulsations, with amplitudes of order 10\%.
For some luminosities, we report similar amplitudes (Fig. \ref{rvst_1d}), but just as
many models have large amplitudes of order unity.
Furthermore, we find, as do they, that the amplitudes, periods, and $L_{\nu_e}$ are correlated.
\citet{buras06a} simulated the collapse, bounce, and
postbounce phases for the 15 M$_{\sun}$ model using a Boltzmann
transport algorithm in 1D and a ``ray-by-ray'' derivative of the same
transport algorithm for 2D simulations.  Generically, they did not
obtain explosions.  However, for a few runs, they
omitted the velocity-dependent terms in their transport
formulation, and this resulted in explosions.  Explosions occurred
soon after the shock stalled in their 2D simulations, aborting any
obvious oscillations in shock
radius.  On the other hand, like our simulations, accretion of the Si/O interface initiates
pulsations with amplitudes $\sim$25\% in their 1D simulations
oscillations. 
.

In reporting the temporal morphology of shocks in 2D simulations, we decompose the shock position,
$R_{\rm shock}(\theta,t)$, into spherical harmonics:
\begin{equation}
R_{\rm shock}(\theta,t) = \sqrt{4 \pi} \sum_{\ell} a_{\ell}(t) Y_{\ell
  0} \, ,
\end{equation}
where
\begin{equation}
\label{eq:al}
a_{\ell}(t) = \frac{1}{\sqrt{4 \pi}}\int R_{\rm shock}(\theta,t) Y_{\ell 0} d \Omega \, ,
\end{equation}
are the time-dependent coefficients, and the normalization for
$a_{\ell}$  in eq. (\ref{eq:al}) ensures that $a_0$ is the average
shock radius.
As is usual,
$Y_{\ell 0}$ are the spherical harmonics, azimuthal symmetry
dictates $m = 0$, and the spherical harmonics are normalized to satisfy
\begin{equation}
\int Y^*_{\ell 0} Y_{\ell 0} d \Omega = 1 \, .
\end{equation}

\begin{figure}
\epsscale{0.75}
\plotone{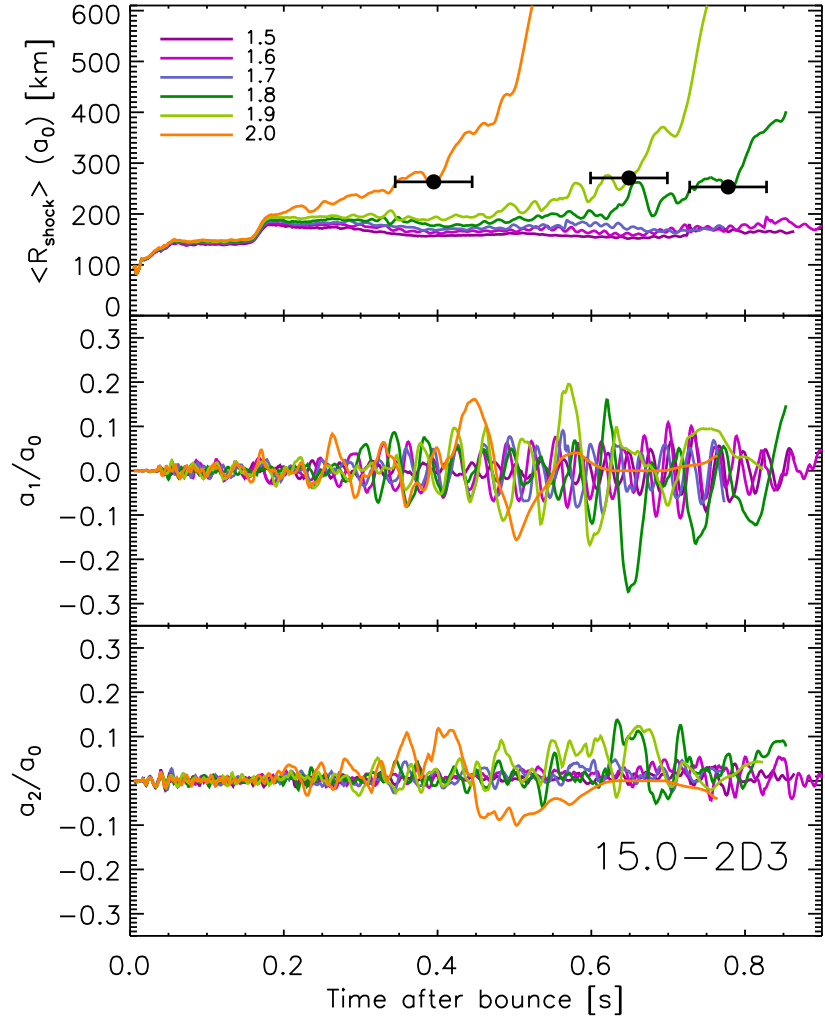}
\caption{Shock morphology vs. postbounce time for the 15.0-2D3
  sequence. The top panel shows $a_0$ or $\langle R_{\rm shock} \rangle$, the average shock
radius, and the middle and bottom panels plot the $\ell = 1$ and
$\ell = 2$ components divided by to the average shock radius:
$a_1/a_0$ and $a_2/a_0$, respectively.  The
filled circles in the top panel indicate $t_{\rm exp}$, the time of explosion, and
the error bars show our estimate of uncertainty ($\pm 50$ ms).
Unlike the 1D models (Fig. \ref{rvst_1d}), the 2D simulations show
very little radial oscillation near the critical luminosity, but
significant nonradial oscillations.  In general, $a_1/a_0$ grows from
0\% to $\sim$10\% for all models, and $a_2/a_0$ grows to $\sim$5\%.  The $a_1/a_0$ amplitudes hint at a correlation with
$L_{\nu_e}$ (see \S \ref{section:shock} for a discussion).  Near
explosion the amplitude of $a_1/a_0$ can peak around $\sim$20\%. \label{sphharm}}
\end{figure}

These temporal coefficients for the
15.0-2D3 sequence are plotted in 
Fig. \ref{sphharm}. The average shock
radius, $a_0$ or $\langle R_{\rm shock} \rangle$, is shown in the top
panel, and a dot indicates the time of explosion, $t_{\rm exp}$ for
the exploding models.  Accompanying are error bars
indicating an estimated uncertainty of $t_{\rm exp}$ (see \S
\ref{section:conditions} for the definition of $t_{\rm exp}$).  The $\ell = 1$ and
$\ell = 2$ components normalized by $a_0$ are shown in the middle and
bottom panels.  First of all, note that the range of luminosities bracketing the critical
luminosity for the 2D
simulations in Fig. \ref{sphharm} are lower than the range for 1D
simulations in Fig. \ref{rvst_1d}.  Further comparison indicates that, unlike 1D simulations, 2D simulations show very little oscillation
in $\langle R_{\rm shock} \rangle$ prior to explosion. In fact, oscillations
in $\langle R_{\rm shock} \rangle$ are miniscule for non-exploding
models and only reach $\sim$10\% around
$t_{\rm exp}$ for exploding simulations.

Instead,
SASI originated oscillations in $\ell=1$ and $\ell=2$ components dominate the non-exploding and exploding models.
The amplitudes of the dipole component, $a_1/a_0$, for $L_{\nu_e}=1.5$
  remain low at $\sim$2.5\%, from bounce until 500 ms.  Around 500 ms, the amplitude 
  grows to $\sim$10\%.
On the other hand, models with $L_{\nu_e} = 1.6$-1.9 show a steady rise to
  $\sim$10\% from bounce.
Exploding models, for which $L_{\nu_e} =1.8$, 1.9, and 2.0, show an abrupt
increase in amplitude to $\sim$20\%.
For non-exploding models, $a_1/a_0$ saturates at $\sim$10\%, and
  for $L_{\nu_e} = 1.5$ and 1.6 it decreases to $\sim$5\%.
At early times, simulations with $L_{\nu_e} = 1.5$ and $L_{\nu_e}
= 2.0$ show the lowest and highest SASI amplitudes, respectively, and are
easily distinguished from the other simulations.  While this would suggest a
correlation in SASI amplitude with $L_{\nu_e}$, the other models do
not show a similar monotonic trend.  Rather their amplitudes occupy
the region between $L_{\nu_e} = 1.5$ and $L_{\nu_e} = 2.0$ but show no
discernible trend with luminosity.
The quadrupole term, $a_2/a_0$, is generally below $\sim$5\% during non-exploding
phases.  For times just before explosion and afterward, $a_2/a_0$
can reach $\sim$10\%.

Two important results are apparent in Figs. \ref{rvst_1d} and
\ref{sphharm}.  A critical luminosity separates exploding and
non-exploding models in time-dependent 1D and 2D simulations, and
the critical luminosity for 2D simulations is $\sim$70\% of the critical luminosity
for 1D models. More evidence will be presented in \S
\ref{section:critical} that the critical luminosity depends upon $\dot{M}$.
For 1D simulations, we observe pulsations near the critical
luminosity, but in 2D simulations, radial oscillations are all but
absent.  Instead, non-radial SASI oscillations dominate.

\section{Effects of the Grid}
\label{section:grid}

The conclusions of the previous section might depend upon the
grid structure used in 2D simulations.
To address this concern, we investigate the dependence of these
results on resolution and the range of polar angle, $\theta$, included
in the simulations.
Encoded in the sequence names (Table \ref{table:sequences}) are the types of grids used for
  2D simulations.  For calculations that encompass the full
  range of $\theta$ from pole to pole (180$^{\circ}$) $\theta$ is divided into 200 zones,
  and for simulations that include only the region between a pole and
  the equator (90$^{\circ}$) $\theta$ is divided into 100 zones.
Sequences that use the 180$^{\circ}$ grid have `2D' in their name,
  and for sequences that use the 90$^{\circ}$ grid, `Q' is in its stead.
The numbers after `2D' or `Q' indicate the resolution.  The
  lowest resolution, denoted by `1', extends the spherical grid from
  the outer edge of the butterfly mesh, 34 km, to 4000 km with 150 radial
  zones.  The middle resolution, `2', divides
  this same space into 300 radial zones, and the highest resolution,
  `3', divides the radii between 34 km and 1000 km into 300 zones.  For all resolutions, the space between the butterfly mesh and
  the outer boundaries are positioned logarithmically.
These simulations do show nontrivial differences in the SASI and
  post-shock flow.
However, the conclusions of the previous section are insensitive to
  changes in resolution and the range of $\theta$ simulated.

\begin{figure}
\epsscale{0.8}
\plotone{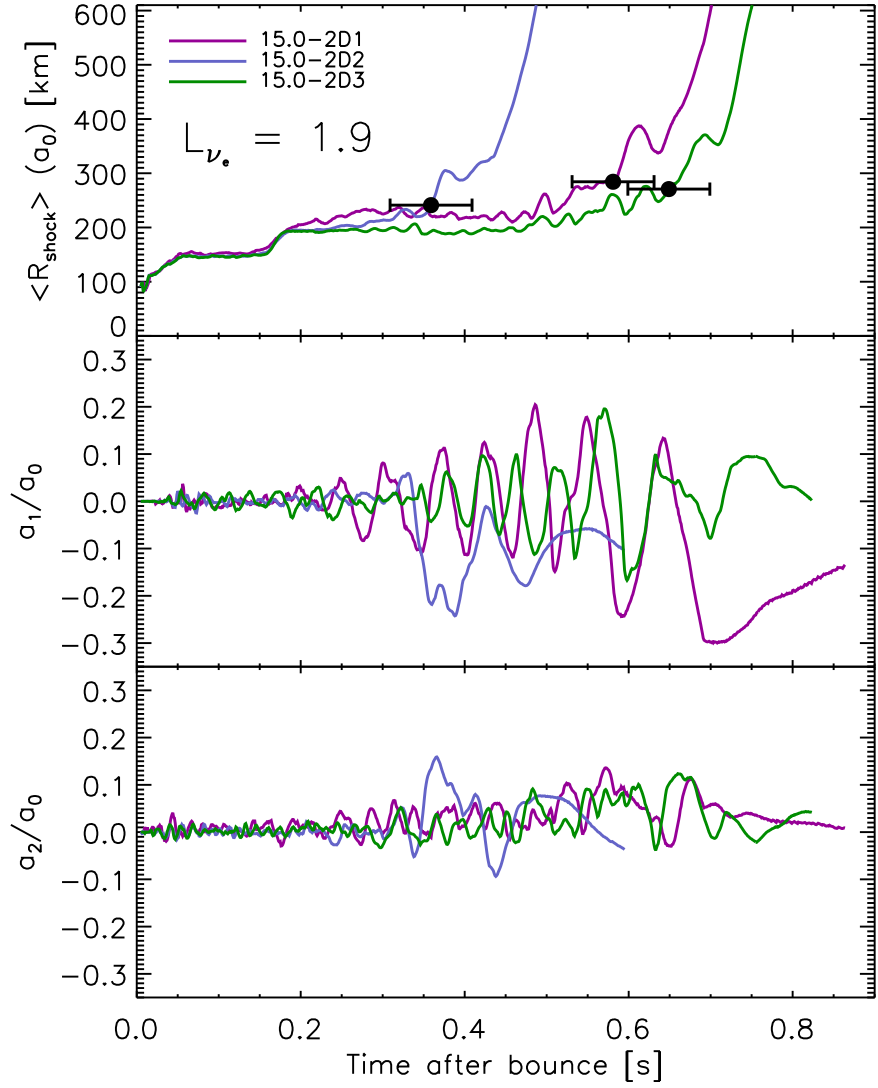}
\caption{The shock morphology vs. time for three
different resolutions of 2D simulations (15.0-2D1, 15.0-2D2, and
15.0-2D3).  The layout is similar to Fig. \ref{sphharm}.  The
electron-neutrino luminosity for each is $1.9 \times 10^{52}$ ergs
s$^{-1}$.  From lowest resolution (15.0-2D1) to the highest resolution
(15.0-2D3), the explosion times are 581, 359, and 649 ms.
Hence, $t_{\rm exp}$ is not monotonic with resolution.  See \S
\ref{section:grid} for a discussion.
\label{morphres}}
\end{figure}

For these three resolutions, Fig. \ref{morphres} shows $\langle R_{\rm shock} \rangle$,
$a_1/a_0$, and $a_2/a_0$ vs. time
for a single luminosity, $L_{\nu_e} = 1.9$, and the 15-M$_{\sun}$ progenitor
model.
Their general evolution is similar to what was discussed in \S
\ref{section:shock} and shown in Fig. \ref{sphharm}.
However, there are some nontrivial differences.

After the Si/O interface is accreted, $\langle R_{\rm
  shock} \rangle$ of the lowest resolution model (15.0-2D1) secularly
creeps out to $\sim$220 km over 100 ms.  Then, outward
progression stalls again, and the model doesn't explode until 581 ms
after bounce.  Both the 15.0-2D2 and 15.0-2D3 models momentarily stall at
$\sim$200 km, a smaller radius. 150 ms later, the average shock radius
of the 15.0-2D2 model reaches
$\sim$220 km on its way to explosion.  At an even later time, 550 ms,
the shock radius of the 15.0-2D3 model reaches $\sim$220 km.  By this
measure, the trend with resolution is monotonic.  However, this does
not translate to monotonicity in explosion time with resolution.
From the lowest resolution (15.0-2D1) to the highest resolution
(15.0-2D3), the explosion times are 581, 359, and 649 ms.
Yet, considering another luminosity, $L_{\nu_e} = 2.0$ (see Table \ref{table:explosions}) does present a monotonic trend toward later
explosion times with higher resolution: from the lowest to the
highest resolution, $t_{\rm exp} = 325$, 364, and
395 ms.  From this, one might conclude that
higher-resolution grids forestall explosion, yet only the
highest resolution shows an explosion for $L_{\nu_e} = 1.8$.

The $\ell = 1$ and $\ell = 2$ (the middle and bottom panels of Fig. \ref{morphres}) spherical
harmonic coefficients, like the monopole term, show some differences
with resolution, but present very few clear trends.  Any trends that do
appear are subtle.  
In the earliest phase, before accretion of the
Si/O interface, the highest resolution
simulations exhibit the greatest amplitude, though
all three resolutions have amplitudes in the $\ell =
1$ coefficient that are less than 5\%.  
At the same time, the $\ell = 2$ component is comparable
for all three resolutions.  After accretion of the Si/O interface,
the nonradial components begin to rise for all three resolutions.
From $\sim$200 to $\sim$325 ms, the 15.0-2D1 model shows the
greatest increase in both nonradial components, while for 15.0-2D2 and
15.0-2D3 they are comparable in growth.  At $\sim$325 ms, the 15.0-2D2
model shows a significant increase in the nonradial components that
coincides with explosion.  The nonradial components for the other two
simulations continue a secular increase in amplitude until their
explosions at later times.  All the while, the lowest resolution run
consistently maintains a slightly larger amplitude in the $\ell = 1$
component, but the $\ell = 2$ components are quite similar.
Generally, before accretion of the Si/O interface, oscillation
amplitudes for $a_1$ are mildly correlated with resolution, and
afterward it is mildly anti-correlated with resolution. 

\begin{figure}
\epsscale{0.75}
\plotone{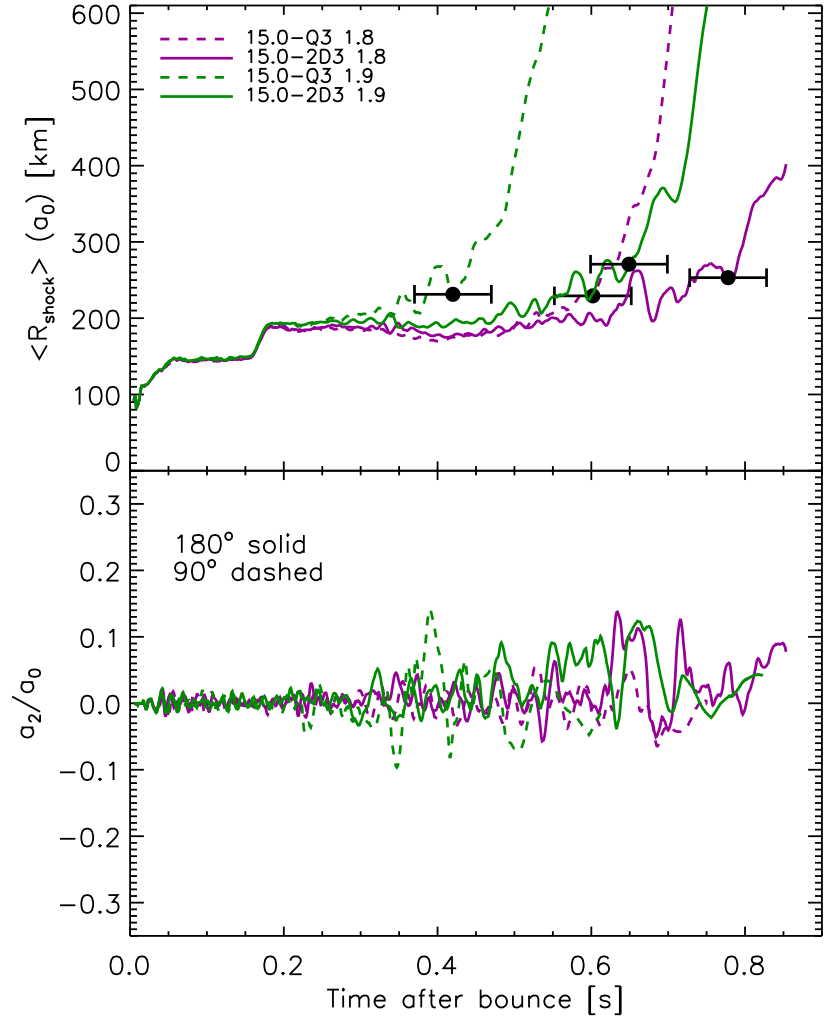}
\caption{$\langle R_{\rm shock} \rangle$ and
$a_2/a_0$ vs. time for two models of 15.0-Q3
and 15.0-2D3.  This compares the differences among the 2D runs using
180$^{\circ}$ (solid) and 90$^{\circ}$ (dashed).  $L_{\nu_e} = 1.8$ corresponds to the
purple curves and $L_{\nu_e} = 1.9$ corresponds to the green curves.
Inspection of $\langle R_{\rm shock} \rangle$ (top panel of
Fig. \ref{fullvs90}) shows that the 90$^{\circ}$ simulations consistently explode
before the 180$^{\circ}$ simulations.  Other than the time near $\sim$300 to $\sim$400 ms, all models show similar evolution in the
amplitude of $a_2/a_0$.  The one exception appears around $\sim$300
to $\sim$400 ms, when the 90$^{\circ}$, $L_{\nu_e} = 1.9$ model explodes and shows the largest amplitude.
\label{fullvs90}}
\end{figure}

Figure \ref{fullvs90} compares 180$^{\circ}$ (15.0-2D3, solid) and 90$^{\circ}$
(15.0-Q3, dashed) simulations.  $\langle R_{\rm shock} \rangle$ and
$a_2/a_0$ vs. time are plotted for two electron-neutrino luminosities $L_{\nu_e} = 1.8$ 
(purple curves) and $L_{\nu_e} = 1.9$ (green curves).  Once again dots
in the top panel mark $t_{\rm exp}$; interestingly, the 90$^{\circ}$
simulations consistently explode before the 180$^{\circ}$ simulations.
Other than for times near $\sim$300 to $\sim$400 ms, all models show similar evolution in the
amplitude of $a_2/a_0$.  The one exception appears around $\sim$300
to $\sim$400 ms, when the 90$^{\circ}$, $L_{\nu_e} = 1.9$ model starts
exploding and shows the largest amplitude.  Even though the 90$^{\circ}$
simulations explode earlier than the 180$^{\circ}$ simulations, the
critical luminosities for the 180$^{\circ}$ and 90$^{\circ}$ simulations
are very similar (see Table \ref{table:explosions} and \S \ref{section:critical}).

Using a ``ray-by-ray'' neutrino transport
    algorithm, \citet{buras06b} performed simulations of the 11.2-M$_{\sun}$ progenitor
    that teetered on the verge of explosion for a 90$^{\circ}$
    calculation, but exploded using 180$^{\circ}$.
They reported no
explosion with a 90$^{\circ}$ simulation but, with all else being equal,
obtained an explosion with a 180$^{\circ}$ simulation.  Differences between our simulations and theirs
include neutrino transport, EOS, and hydrodynamic solver.  While any
of these could account for the apparent discrepancy, we emphasize the
difference in the 90$^{\circ}$ grids.  The 90$^{\circ}$ simulations of
\citet{buras06b} included a wedge that extends plus and minus 45$^{\circ}$ of
the equator and suppressed $\ell = 1$ and $\ell = 2$ modes, the two
prominent modes of the SASI.  Our 90$^{\circ}$ simulations range from the pole to the equator
and inhibit only the $\ell = 1$ mode.  Regardless, \citet{buras06b} noted that
their 90$^{\circ}$ simulation was close to explosion and that extending
the grid to 180$^{\circ}$ simply tipped the scales toward explosion.
Using the same
hydrodynamics and neutrino transport as \citet{buras06b}, but a
90$^{\circ}$ grid that goes from the pole to the equator,
\citet{marek07b} found that 90$^{\circ}$ simulations explode
insignificantly earlier than 180$^{\circ}$ simulations.
In either case, these results are consistent with our finding that the
critical luminosity for simulations using 90$^{\circ}$ and
180$^{\circ}$ are very similar.

Resolution and the angular size of the grid non-trivially affect the shock
position and morphology in 2D simulations.
For one, the lowest resolution simulations
consistently maintain a slightly larger amplitude in the $\ell = 1$
component.  At early times, the lowest resolution simulations reach
$\sim$220 km before higher resolution simulations.  This does not lead to a
monotonic correlation in explosion times, though.
In changing the angular size of the grid, the 90$^{\circ}$ simulations
consistently explode at earlier times compared to the 180$^{\circ}$
simulations.
Despite differences that result from grid changes, the
conclusions of the previous section do not change.  Specifically, the
critical luminosity for the 90$^{\circ}$ and 180$^{\circ}$ simulations are
very similar.

\section{Heating and Advection Timescales}
\label{section:timescales}

Measures that are related to the critical $L_{\nu_e}$ and $\dot{M}$
condition \citep{burrows93} are the
  timescale for matter to traverse the gain region, the advection
  timescale, $\tau_{\rm adv}$, and the heating timescale of the matter
  in the gain region, $\tau_q$.
\citet{thompson00} argued that 1D core-collapse simulations do not
explode because $\tau_q$ is longer than $\tau_{\rm adv}$.  In other
words, core-collapse simulations fail because $\tau_{\rm adv} / \tau_{q}
< 1$.  He defined, $\tau_q \sim aT^4/\dot{Q}$, and
    the advection time as the flow time across a scale height.
\citet{janka01} explored similar integral concepts, and proposed
conditions for explosion that are akin to
\begin{equation}
\label{eq:taucondition}
\tau_{\rm adv} / \tau_q > 1 \, ,
\end{equation}
the opposite of the \citet{thompson00} failure condition.
\citet{thompson05} applied this
    condition to 1D core-collapse
    simulations and demonstrated that this ratio does indeed help to
    distinguish non-exploding models from exploding models.  They
    defined the advection timescale as $H/v_r$, where H is the
    pressure scale height, and the heating timescale as
    $(P/\rho)/\dot{q}$, where $\dot{q}$ is the local net
    heating rate.
Since, others have explored $\tau_{\rm adv}$ and $\tau_q$ as a condition
for explosion in 2D simulations \citep{buras06a,scheck08}.
While $\tau_{\rm adv}/\tau_{q} >
1$ is to some extent useful in identifying explosions, our results suggest
that it is only useful as a rough diagnostic.

We define the heating time as a characteristic time for
  neutrinos to change the thermal energy of the matter in the gain
  region, where heating dominates cooling,
  \begin{equation}
    \label{eq:tauq}
    \tau_q = \frac{\int_{\rm gain} (\varepsilon - \varepsilon_0) dm }
    {\dot{Q}} \, ,
  \end{equation}
  where $\varepsilon_0$ is the zero point energy for
  the EOS and 
  \begin{equation}
    \label{eq:qdot}
    \dot{Q} = \int_{\rm gain} (\mathcal{H} - \mathcal{C}) dm \,
  \end{equation}
  is the net heating rate and is integrated over the gain region.
While \citet{scheck08} and \citet{buras06b} include $1/2v^2 + \Phi$ in the
    integrand of eq. (\ref{eq:tauq}), we suggest,
    like \citet{thompson00} and \citet{thompson05}, that comparing the
    heating rate to the thermal energy eq. (\ref{eq:tauq}) is a more natural definition of a heating
    timescale. To be clear though, these differences in definition
    alter the absolute scale, but not the general trends.

For spherically symmetric flow, the advection timescale is 
  \begin{equation}
    \label{eq:tauadv}
    \tau_{\rm adv} = \int_{\rm gain} \frac{dr}{v_r} \,
  \end{equation}
  where the the integral is integrated from the shock position to the
gain radius.
For 2D simulations, many incarnations of eq. (\ref{eq:tauadv}) have
appeared in the literature.  \citet{scheck08} use a solid-angle
averaged version of eq. (\ref{eq:tauadv}) for $\tau_{\rm adv}$.
\citet{buras06b} use eq. (\ref{eq:tauadv}) for the advection
    timescale in 1D, but for 2D simulations they redefine this
    timescale as the time for a mass shell to traverse the gain region.
We've explored three measures of the advection timescale.
The first replaces the denominator in eq. (\ref{eq:tauadv})
    with the solid-angle averaged velocity, $\langle v_r \rangle =
    (\int{v_r d \Omega})/4 \pi$.
For the second, if one assumes steady-state flow, then
    $\tau_m = M_{\rm gain}/\dot{M}$ is a useful measure of the
    accretion timescale, where $M_{\rm gain}$ is the mass in the gain
    region and $\dot{M}$ is the mass flux in or out of the gain
    region.  For simplicity, we use $\dot{M}$ just exterior to the
    shock.  The advantage of this definition is that it is relatively
    straightforward to measure $M_{\rm gain}$ and $\dot{M}$ for 2D simulations.
Prior to explosion, these two definitions give similar results, but
as the model explodes, the assumption of steady-state accretion is violated
and $\tau_m$ begins to underestimate $\tau_{\rm adv}$.
Neither of these definitions, however, fully capture the complex
    multi-dimensional nature of the post-shock flow.

We define a new timescale that better captures the multi-dimensional
    nature of the post-shock flow.
Post-processing the simulation data, we integrate the paths of tracer
particles for 150 ms, record their
    trajectories, and define a residence time, $\tau_r$, which is the
    duration of each particle's time spent in the gain region.
50,000 particles are initiated at either 400 km, a
    radius well outside the shock, or 150 km, the middle of the gain
    region.
At either radius, the particles are randomly distributed in $\mu = \cos \theta$.
For most situations,
the particles traverse the gain region within the integration time,
150 ms, and have
    accreted onto the PNS.
Every 10 ms, a new generation of 50,000 particles is generated,
providing a time-dependent distribution of $\tau_r$.

\begin{figure}
\epsscale{0.7}
\plotone{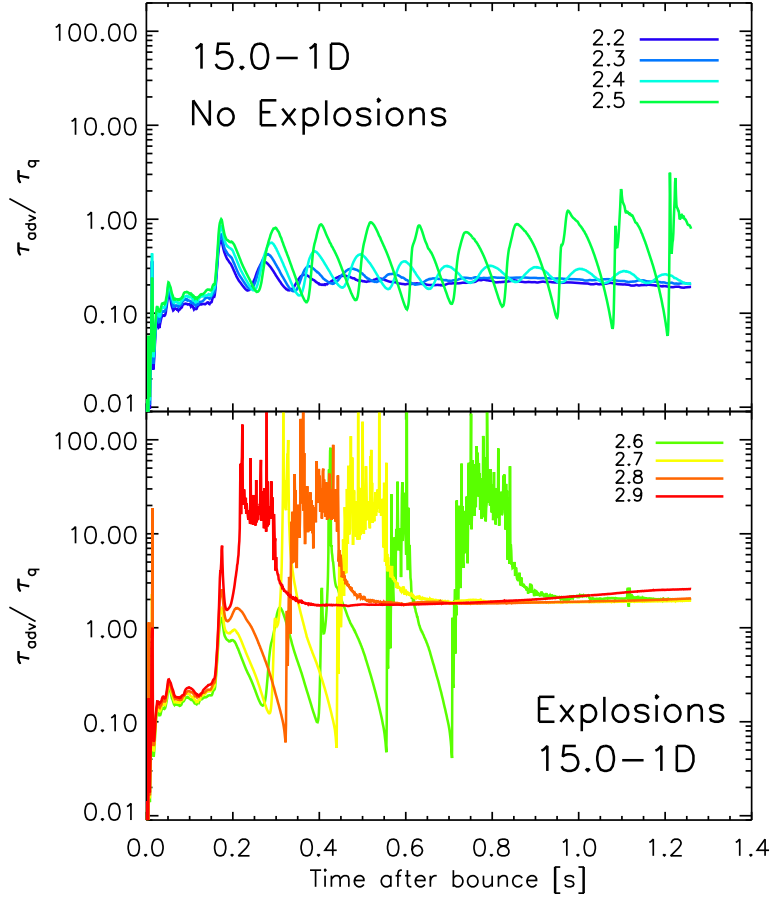}
\caption{The ratio of advection and heating timescales, $\tau_{\rm adv}/\tau_{\rm
  q}$, vs. postbounce time for the 15.0-1D sequence.  The top
panel shows the models that do not explode by 1.3 s after bounce, and
the bottom panel shows those that do explode.  Despite dramatic
oscillations, $\tau_{\rm adv}/\tau_{\rm q}$ rarely reaches a value of
above $\sim$1 for the non-exploding models (top panel).  In all non-exploding
cases, except $L_{\nu_e} = 2.5$ (in units of $10^{52}$ erg s$^{-1}$), the
oscillations decay (see text and the caption of Fig. \ref{rvst_1d} for
the timescales).  For exploding
models (bottom panel), $\tau_{\rm adv}/\tau_{q}$ makes large
excursions from $\sim$0.1 to $\sim$100.  Larger luminosities produce more excursions,
with one at $L_{\nu_e} = 2.9$ and four at
$L_{\nu_e} = 2.9$.  We conclude that $\tau_{\rm
  adv}/\tau_{q} > 1$ is a useful diagnostic, but not a rigorous condition for explosion.\label{tscalesvst_1d}}
\end{figure}

The ratio $\tau_{\rm adv}/\tau_q$ vs. time is shown in Fig. \ref{tscalesvst_1d} for the 15.0-1D sequence.  The top
panel shows the models that do not explode by 1.3 s after bounce, and
the bottom panel shows those that do explode.  Despite dramatic
oscillations, $\tau_{\rm adv}/\tau_{\rm q}$ rarely reaches a value of
one for the non-exploding models (top panel).  In all non-exploding
cases, except $L_{\nu_e} = 2.5$, the
oscillations in this ratio decay with a timescale commensurate with the decay times
of the shock radius oscillations (Fig. \ref{rvst_1d}).  For exploding
models (bottom panel), $\tau_{\rm adv}/\tau_{q}$ executes large
excursions from $\sim$0.1 to $\sim$100.  Note that $\tau_{\rm
  adv}/\tau_{\rm q}$ makes several excursions to values above order
unity and back again.  Smaller luminosities produce more excursions,
with one at $L_{\nu_e} = 2.9$ and four for
$L_{\nu_e} = 2.6$.

\begin{figure}
\epsscale{0.75}
\plotone{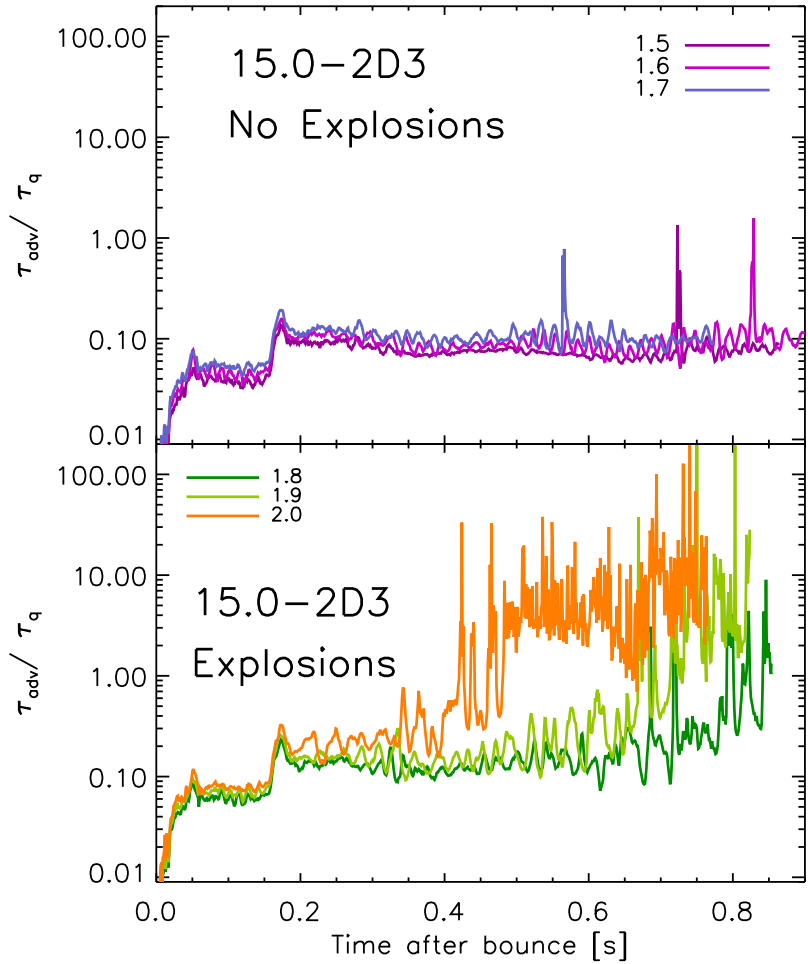}
\caption{Similar to Fig. \ref{tscalesvst_1d}, but we plot
$\tau_{\rm adv}/\tau_{q}$ vs. postbounce time for the 15.0-2D3 sequence.  The models that don't explode, $L_{\nu_e} =
1.5$, 1.6, and 1.7 (in units of $10^{52}$ erg s$^{-1}$) show relatively constant
ratios for all times.  On the other hand, the exploding models,
$L_{\nu_e} = 1.8$, 1.9, and 2.0, start at low ratios ($\sim$0.1)
and increase to ratios of order $\sim$10.  Unlike the 1D
models, once this ratio reaches large values, it
remains there for the rest of the simulation.  The times of explosion
are 395, 648, and 778 ms for $L_{\nu_e} = 1.8$, 1.9, and 2.0,
respectively.  Notice that the ratio of timescales begin to rise
before $\tau_{\rm adv}/\tau_q = 1$, suggesting that these models are
on their way to explosion before the condition $\tau_{\rm adv}/\tau_q
> 1$ is met.\label{tscalesvst_2d}}
\end{figure}

The ratio of these timescales for 2D simulations (15.0-2D3 sequence) are shown
in Fig. \ref{tscalesvst_2d}.  The models that don't explode, $L_{\nu_e} =
1.5$, 1.6, and 1.7, show relatively constant
ratios for all time.  On the other hand, the exploding models,
$L_{\nu_e} = 1.8$, 1.9, and 2.0, start at a low ratio ($\sim$0.1)
and secularly increase to ratios of order $\sim$10.  Unlike the 1D
models, once this ratio reaches large values (in an average sense), it
remains there for the rest of the simulation.

Generally, the condition $\tau_{\rm
  adv}/\tau_{q} > 1$ is a useful diagnostic of explosion, but
given the ambiguities in defining
    $\tau_{\rm adv}$ and $\tau_q$, its accuracy is limited. 
For example, 1D simulations that explode make dramatic excursions above and
    below this line before explosion.
It
might be more sensible to state that this condition should be
satisfied for a finite time before explosion occurs.  
However, one shouldn't expect rigor when defining a heating timescale.
Note that this condition assumes a quasi-steady state, and the
large amplitude oscillations in 1D complicate this assumption.
Finally, $\tau_{\rm adv}/\tau_q$, in 2D exploding simulations, increases
monotonically for $\sim$100 ms, while this ratio remains low for 2D
non-exploding simulations.  This distinction implies that
2D exploding models are well on their way to explosion before
$\tau_{\rm adv}/\tau_q > 1$ is satisfied.  Hence, $\tau_{\rm adv}/\tau_q$
is a useful diagnostic for explosion, but may not be a rigorous
condition for explosion.

In Fig. \ref{taures_mean_frac}, we plot for all generations of tracer
particles the mean residence times, $\langle
\tau_r \rangle$ (top panel), and the fraction ($f_r$) of particles with
$\tau_r > 40$ ms (bottom panel) as a function of
time for the $L_{\nu_e} = 1.9$ model of the 15.0-2D3 sequence.  The
particles with $R_{\rm start} = 400$ km (150 km) are shown with solid
(dashed) lines.  Both $\langle \tau_r \rangle$ and $f_r$ show that the particles with
$R_{\rm start} = 400$ km and 150 km track two fundamentally
different flows.  Generally, the particles with
$R_{\rm start} = 150$ km have longer average residence times and more
particles with $\tau_r > 40$ ms than the generation of particles with
$R_{\rm start} = 400$ km, which suggests that most of the mass
that accretes through the shock quickly finds its way to the PNS through narrow, low entropy plumes.  Despite the quick
accretion of new matter, the long $\tau_r$ of particles with $R_{\rm
  start} = 150$ km indicate that a significant amount of mass in the
gain region does linger, resulting in more heating and higher
entropies.

To verify this, we compute the total
amount of mass with high entropies and compare
this to the total mass in the gain region.  To define high entropies,
we choose a minimum entropy of 14 k$_{\rm B}$/baryon, which corresponds
to the maximum entropy for the 1D simulation of the same neutrino
luminosity (see Fig. \ref{entropy}) and gives a rough boundary between
the regions with short (low entropy) and long (high entropy) residence
times.  The ratio of mass with high entropy to the total mass in the gain region
grows from 0\% to 70\% between 0.2 s and 0.5 s after bounce.  For the
rest of the simulation the ratio remains at this level.  Therefore, even though
most of the accreted mass advects quickly through the gain region via the
down-flowing plumes, a significant fraction of the mass in the gain
region is better characterized with long residence times and high entropies.

In addition, $\langle \tau_r \rangle$ and
$f_r$ show four distinct phases, with each successive phase
having longer $\langle \tau_r \rangle$ and higher $f_r$.
Distributions of $\tau_r$ that represent these four phases are shown in
Fig. \ref{taures}, and spatial colormaps of entropy are shown in
Figs. \ref{stills1}, \ref{stills2}, and \ref{stills3} that
correspond to the beginning (Fig. \ref{stills1}), middle
(Fig. \ref{stills2}), and end (Fig. \ref{stills3}) of each generation's
integration time of 150 ms.    The top panel of
Fig. \ref{taures} shows the
distribution for four generations with $R_{\rm start} = 400$ km and at $t = 0.130$, 0.370, 0.530, and 0.6 s after
bounce.  The bottom panel shows similar residence-time distributions
for generations with $R_{\rm start} = 150$ km.

The entropy maps in the top-left panels of Figs. \ref{stills1}, \ref{stills2}, and \ref{stills3} show that there
is very little convection or SASI for the first phase ($t = 0.130$ s),
and the distributions of $\tau_r$ (Fig. \ref{taures}) show that all particles traverse
the gain region quickly, within $\sim$9 ms.  At $\sim$150 ms, when the Si/O interface
is accreted, both $\langle \tau_r \rangle$
and $f_r$ show a dramatic
increase.  The mean residence time increases from $\sim$10 ms to
$\sim$40 ms for $R_{\rm start}=400$ and 150 km, and $f_r$ increases
from 0\% to 35\% for $R_{\rm start} = 400$ km and $\sim$50\% for $R_{\rm
  start} = 150$ km.  The second phase, from
$\sim$200 ms to $\sim$450 ms, corresponds
to the growth and nonlinear saturation of the SASI (top-right panels of Fig. \ref{stills1}-\ref{stills3}).  During this phase ($t = 0.370$ s in Fig. \ref{taures}), $\langle \tau_r
\rangle$ for $R_{\rm start} = 400$ km (150 km) is $\sim$22 ($\sim$32), and $f_r$ is $\sim$10\% ($\sim$30\%).  The third phase, from $\sim$0.45 to $\sim$0.6 s,
shows a rise in both $\langle \tau_r \rangle$ and
$f_r$.  The distributions in Fig. \ref{taures} that represent this phase, $t= 0.530$ s, have mean residence times of $\sim$50 ms ($\sim$30 ms) for $R_{\rm start} = 150$ km (400 km) and $f_r$ of 50\% (20\%).  The fourth and final phase is marked by a significant increase of
$\langle \tau_{\rm res} \rangle$ and $f_{\rm res}$ for $R_{\rm start}=
150$ km particles and continued rise for $R_{\rm start}= 400$ km.  In
Fig. \ref{taures}, the example distribution is shown at $t =
0.6$ s.  For $R_{\rm start}= 400$ km, $\langle \tau_{\rm res} \rangle \sim 40$ ms
and $f_{\rm res} \sim 25$\%, while $\langle \tau_{\rm res} \rangle \sim 90$ ms
and $f_{\rm res} \sim 70$\%.  This phase is also characterized by a
large bump of particles with very long residence times in
Fig. \ref{taures}, which is much larger for $R_{\rm start}=150$ km than
for $R_{\rm start} = 400$ km.  The appearance of the bump at large $\tau_r$ is one of the most pronounced indicators that the star has reached an explosive situation.

In Fig. \ref{entropy}, we compare 1D entropy profiles of the 15.0-1D,
$L_{\nu_e} = 1.9$ simulation (solid red lines) with the entropy
distributions shown in Fig. \ref{stills1}.  For each zone of the 2D
simulation, points mark the entropy and radius from the center.
Generally, both 1D and 2D profiles show the shock as an abrupt jump in
entropy of the accreting matter, the gain region as
indicated by the highest entropies, and the
cooling region where entropy is significantly reduced as matter continues to accrete onto the PNS.  Where 1D
profiles show a negative entropy
gradient, 2D simulations are convective.  The most striking
differences, though, are that when the SASI dominates the post-shock flow, $t=
0.370$, 0.530, and 0.600 s, the shock is asymmetric, and the gain region has a large range of
entropies, in which high entropies correspond
to regions of long residence times, and low entropies
correspond to plumes that funnel matter toward the PNS.

Other than passage through shocks, the entropy is determined by
neutrino heating and cooling.  Therefore, entropy is a good
measure of the integrated history of net heating.  Considering mass- and volume-weighted averages, the entropy
for 2D simulations is a couple $k_B$/baryon (less than 10\%) higher than 1D
simulations.  However, Fig. \ref{entropy} shows that 2D simulations have
regions with $\sim$30\% higher entropies than 1D simulations for the
same $L_{\nu_e}$ and $\dot{M}$.  In other words, the long residence
times that accompany the SASI in 2D simulations result in regions
with higher integrated net heating and offer an explanation for the
lower critical luminosity of 2D simulations.

\section{Conditions at Explosion}
\label{section:conditions}

As relevant indicators of the core-collapse mechanism, we
investigate the net heating rate ($\dot{Q}$), heating efficiency
($\dot{Q}/L_{\nu_e \bar{\nu}_e}$), and the mass in the gain region ($M_{\rm gain}$).  In general, these quantities show distinctly
different evolutions for 1D and 2D simulations near their respective
critical luminosities.  Plots of the 1D simulations reflect the very
  large radial shock oscillations, while plots of the 2D simulations
  show relatively smooth evolution and bifurcations distinguishing
  explosive models from non-explosive models.  
\citet{scheck08} compared these quantities for 1D and 2D simulations
  with the same neutrino luminosity and found similar bifurcations in
  these integral measures.  We show that this bifurcation is generic
  when we include 1D and 2D non-exploding and 2D exploding models. 
In Table \ref{table:explosions}, we summarize the conditions at
explosion, but first we take care to define $t_{\rm exp}$, the time of
explosion.

The time of explosion for 1D
simulations is quite pronounced and occurs near the time when
the ratio $\tau_{\rm adv}/\tau_q$ becomes of order unity
or larger for the last time.  However, for 2D
simulations, all measures show a gradual trend toward explosion
over 50 to 100 ms timescales.  Therefore, to define $t_{\rm exp}$
we identify the last time that $\tau_{\rm adv}$ is less than 50 ms (65
ms) for 15-M$_{\sun}$ (11.2-M$_{\sun}$) models.
This definition establishes the time of explosion after there is a
clear trend toward explosion, but before post-shock material is
launched outward.  Acknowledging the ambiguity of such a definition, we assign an uncertainty to $t_{\rm exp}$ of $\pm 50$
ms.

\begin{figure}
\epsscale{0.7}
\plotone{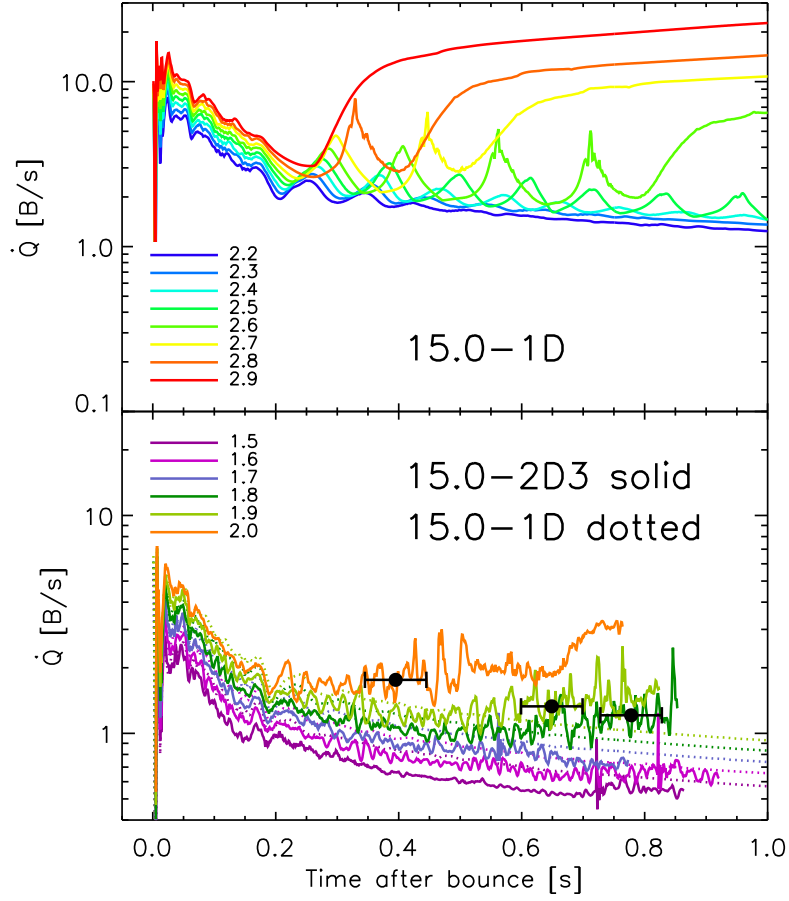}
\caption{Time evolution of the net heating rate, $\dot{Q}$, for
the 15.0-1D and 15.0-2D3 sequences.
The top panel plots $\dot{Q}$ for the 1D sequence only for the range of $L_{\nu_e}$
near the 1D critical luminosity.  The bottom panel highlights the range
of $L_{\nu_e}$ near the critical luminosity of the 2D sequence, and
plots both 1D and 2D simulations.  The dots show the $\dot{Q}$ at the time of
explosion for the 2D simulations.  $\dot{Q}$ shows pronounced
oscillations for the 1D simulations near explosion (top panel), with the
peaks corresponding to dips in shock radius in Fig. \ref{rvst_1d}.
For non-exploding 1D and 2D simulations
$\dot{Q}$ evolves downward.
Exploding 2D simulations, however, have $\dot{Q}$ that tends toward
flat or upward evolutions.
\label{qdotvstime}}
\end{figure}

\begin{figure}
\epsscale{0.8}
\plotone{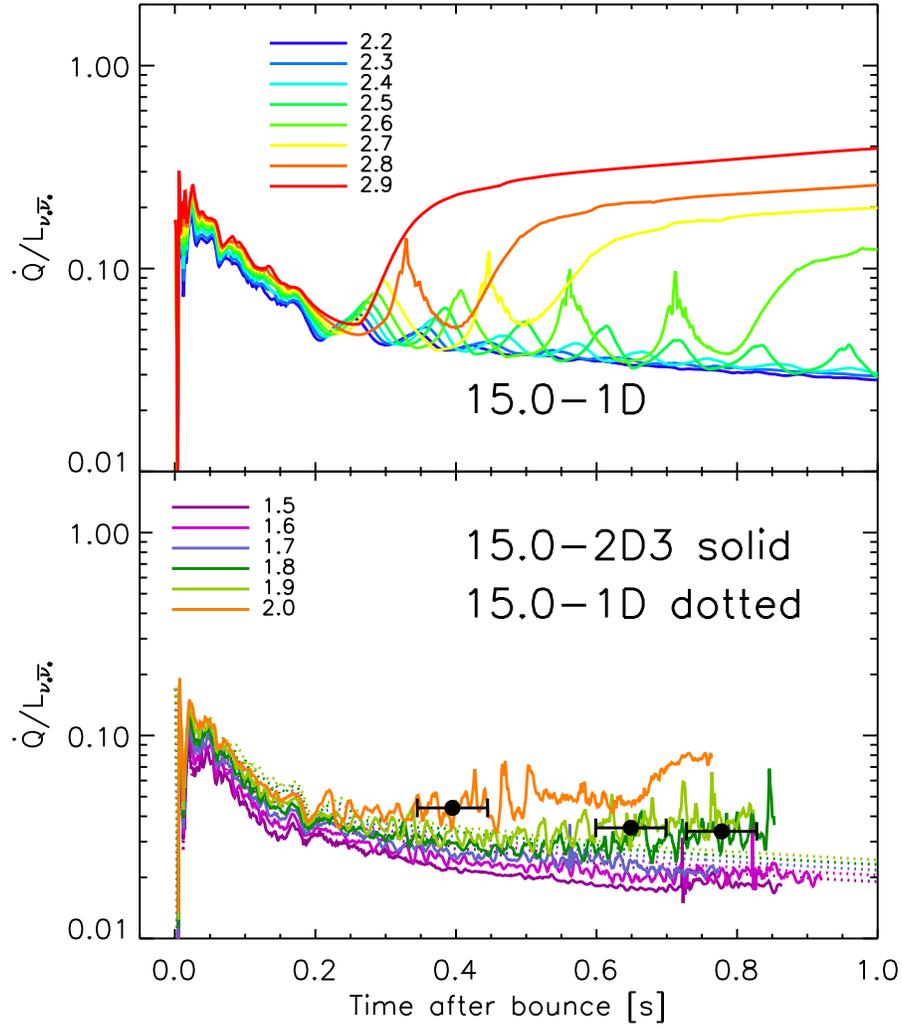}
\caption{Similar to Fig. \ref{qdotvstime}, but here we plot the heating
  efficiency, $\dot{Q}/L_{\nu_e \bar{\nu}_e}$, vs. postbounce time.  The
efficiencies at $t_{\rm exp}$ for the 1D sequence near the critical
luminosity range from 3\% to
10\%, while the efficiencies of the 2D simulations at $t_{\rm exp}$ are
3.4\%, 3.5\%, and 4.4\% for $L_{\nu_e} = 1.8$, 1.9, and 2.0,
respectively.  For non-exploding 1D and 2D simulations
$\dot{Q}/L_{\nu_e \bar{\nu}_e}$ evolves downward.
Exploding 2D simulations, however, have efficiencies that tend to stay
flat or increase.  \label{effvstime}}
\end{figure}

\begin{figure}
\epsscale{0.8}
\plotone{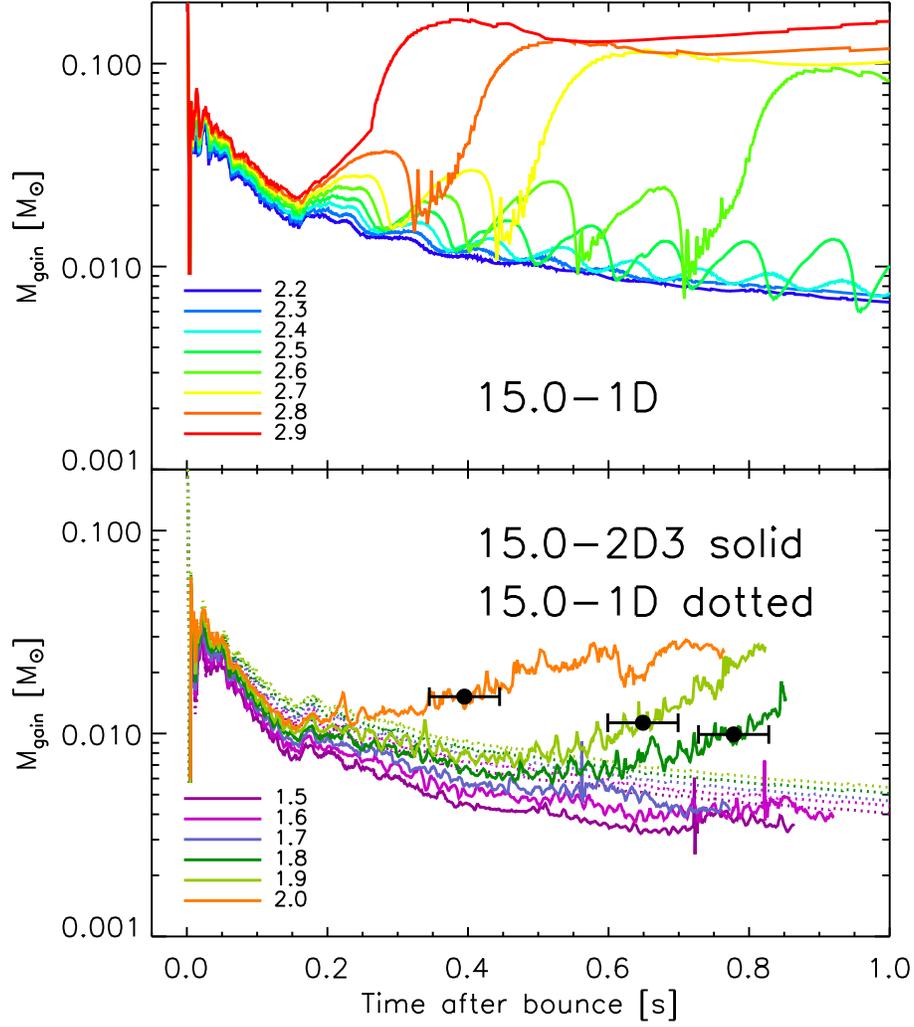}
\caption{Similar to Fig. \ref{qdotvstime}, but here we plot the mass in
  the gain region, $M_{gain}$, vs. postbounce time.  The oscillations
  of $M_{\rm gain}$ in the 1D models (top panel) correspond to the
  oscillations in shock radius.  The minima in $M_{\rm gain }$ correspond to the minima
in shock radius.  In the exploding models of 15.0-2D3, there is a
secular increase in the amount of mass in the gain region leading up
to and past explosion that is distinct from the non-exploding 1D and
2D models. \label{mass_gain}}
\end{figure}

Having defined $t_{\rm exp}$, we now define the conditions at explosion.
For 1D simulations, $\dot{M}_{\rm exp}$ is defined at 250 km
and at $t_{\rm exp}$.  For 2D simulations, $\dot{M}_{\rm exp}$ is
defined just exterior to the shock (at roughly 300 to 250 km) at the
time of explosion.  Corresponding to the uncertainty in $t_{\rm
  exp}$, we determine the uncertainty in $\dot{M}_{\rm exp}$.
Figs. \ref{qdotvstime}, \ref{effvstime}, and \ref{mass_gain}, show a
general upward trend of $\dot{Q}$, $\dot{Q}/L_{\nu \bar{\nu}_e}$, and
$M_{\rm gain}$ around
the time of explosion.  The values in Table \ref{table:explosions}
are obtained by averaging $\dot{Q}$, $\dot{Q}/L_{\nu \bar{\nu}_e}$, and
$M_{\rm gain}$ for a 50 ms range centered on $t_{\rm exp}$.  We
also experimented with a range of 10 and 100 ms and obtained similar
results.  For 1D simulations, we note the conditions at the time of
explosion, but due to large nonlinear oscillations the specific values
are sensitive to the definition of $t_{\rm exp}$.

Figure \ref{qdotvstime}
compares the time evolutions of the net heating rates, $\dot{Q}$, for
the 15.0-1D and 15.0-2D3 sequences.  
While the top panel shows $\dot{Q}$ for the
sequence of 1D simulations near explosion, the bottom panel shows $\dot{Q}$ for the sequence of 2D simulations (solid) near
explosion.  For comparison, 1D models (dotted) for the
same range of $L_{\nu_e}$ as the 2D sequence are included.  The dots show $\dot{Q}$ at the time of
explosion for the 2D simulations.
As expected, $\dot{Q}$ correlates with $L_{\nu_e}$
(Fig. \ref{qdotvstime}).  Consequently, the range of $\dot{Q}$ for the
1D exploding models ($\sim$1 to $\sim$11 B/s, where 1 Bethe (B) $=
10^{51}$ erg) is higher than the
range of $\dot{Q}$ for the 2D exploding models ($\sim$0.5 to $\sim$6
B/s).  The 1D sequence in the top panel shows pronounced oscillations
in $\dot{Q}$ that correspond to the oscillations in shock radius in
Fig. \ref{rvst_1d}.  The maxima in $\dot{Q}$ coincide with the minima
in the shock radius, which has been observed before for 1D simulations
\citep{buras06a}.

The 2D sequence, on the other hand, shows no
such oscillations in $\dot{Q}$.  Rather, the bottom panel of Fig. \ref{qdotvstime}
shows a relatively smooth evolution with time.  After bounce,
$\dot{Q}$ evolves secularly downward for non-exploding 1D and
2D simulations.  In contrast, 50 ms to 100 ms before $t_{\rm exp}$,
$\dot{Q}$ for exploding 2D simulations remains flat with time or even
begins to rise slightly.  In either case, there is a clear bifurcation
in the evolution for exploding 2D simulations and non-exploding 1D and
2D simulations.

Similar to Fig. \ref{qdotvstime}, in Fig. \ref{effvstime} we plot the heating efficiency,
$\dot{Q}/L_{\nu_e \bar{\nu}_e}$.  Even after dividing by $L_{\nu_e \bar{\nu}_e}$
the efficiency has a positive correlation with $L_{\nu_e}$.  This
suggests that the correlation between $\dot{Q}$ and $L_{\nu_e}$ is
steeper than linear.  The
efficiencies at $t_{\rm exp}$ for the 1D sequence range from 3\% to
10\%, while the efficiencies of the 2D simulations at $t_{\rm exp}$ are
3.4\%, 3.5\%, and 4.4\% for $L_{\nu_e} = 1.8$, 1.9, and 2.0, respectively. 
Naturally, the trends in $\dot{Q}/L_{\nu_e
  \bar{\nu}_e}$ are similar to the those of $\dot{Q}$, including the
bifurcation of $\dot{Q}/L_{\nu_e \bar{\nu}_e}$ for exploding and
non-exploding models.

Fig. \ref{mass_gain} shows the time evolution of $M_{\rm gain}$ for
the 15.0-1D and 15.0-2D3 sequences.  Once again, the oscillations in
$M_{\rm gain}$ correspond to the oscillations in shock radius, but
this time the minima in $M_{\rm gain }$ correspond to the minima
in shock radius.  In the exploding models of 15.0-2D3, there is a clear
secular increase in the amount of mass in the gain region leading up
to and past explosion.  This is consistent with the idea that the SASI
helps to increase the amount of mass in the gain region, tipping the
scales toward explosive solutions.  However, we note in
Fig. \ref{sphharm} that even the non-exploding models show some $\ell =
1$ oscillations, and hence, the SASI is necessary but not
sufficient in increasing the mass in the gain region.  At
$t_{\rm exp}$, $M_{\rm gain} =$ 0.099, 0.0113, and 0.0152 M$_{\sun}$ for
$L_{\nu_e} = 1.8$, 1.9, and 2.0, respectively.

We summarize the conditions at explosion in Table
\ref{table:explosions}.  We list for each sequence in Table
\ref{table:sequences}, the electron-neutrino
luminosity, $L_{\nu_e}$ (column 1), time of explosion, $t_{\rm exp}$
(column 2), accretion rate at the time of explosion near the
shock, $\dot{M}_{\rm exp}$ (column 3), the net heating rate, $\dot{Q}$
(column 4), the heating efficiency, $\dot{Q}/L_{\nu_e
  \bar{\nu}_e}$ (column 5), and the mass in the gain region at the
time of explosion, $M_{\rm gain}$ (column 6).  The mass accretion rate
at $t_{\rm exp}$ is addressed in the next section.  The 11.2-1D sequence
manifests a wide range of $\dot{Q}$, from 0.63 to 4.46 B/s, and the range
of the corresponding heating efficiencies is similarly wide, from 2.4\% to 14\%.  In contrast,
the 15.0-1D models have much larger $\dot{Q}$ at $t_{\rm exp}$, but
with a smaller range.  Typically they are $\sim$3.5 B/s with one at
5.57 B/s.  The heating efficiencies are similarly high, $\sim$5.9 to
$\sim$10\%.  On the other hand, 2D simulations have lower efficiencies
at explosion.  All 2D
models that use the 15-M$_{\sun}$ progenitor have $\dot{Q}$ that lies
in the range $\sim$1 to $\sim$2 B/s, and have efficiencies that
range from $\sim$3 to $\sim$4\%.  For most models, $M_{\rm gain}$ at $t_{\rm
  exp}$ is $\sim$0.01 M$_{\sun}$, but can be as low as 0.0073 M$_{\sun}$ and
as high as 0.0319 M$_{\sun}$. 

In Figs. \ref{qdotvstime}, \ref{effvstime}, and
\ref{mass_gain}, the exploding 2D models show distinctly different
values of these indicators
compared to those of the non-exploding 1D and 2D simulations.  $\dot{Q}$,
$\dot{Q}/L_{\nu_e \bar{\nu}_e}$, and $M_{\rm gain}$ trend downward for
the non-exploding models.  Just before, during, and after explosion,
the heating rate, heating efficiency, and gain mass increase with time
for the 2D exploding models.
\citet{scheck08} have shown a similar trend in comparing 1D and 2D
models at the same neutrino luminosity, and we show that this
bifurcation exists among the 2D exploding and non-exploding models as well.

\section{Critical Luminosity and Mass Accretion Rate}
\label{section:critical}

\begin{figure}
\plotone{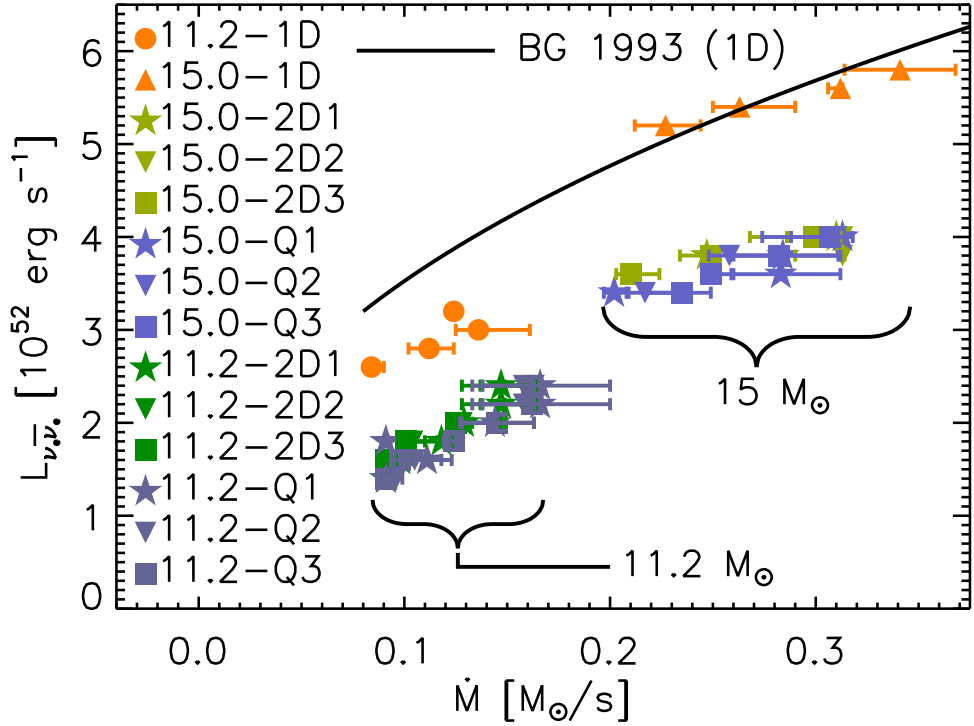}
\caption{The electron- and anti-electron-neutrino luminosity, $L_{\nu_e \bar{\nu}_e}$ ,
  vs. mass accretion rate, $\dot{M}$, of explosion.  The accretion rate is defined just
  exterior to the shock at $t_{\rm exp}$.  Because the time of
  explosion is difficult to define, we also show the range of
  accretion rates 50 ms before and after $t_{\rm exp}$
  as error bars.  In general, 1D simulations are represented by
  orange, 2D simulations by green hues, and 2D-90$^{\circ}$ runs by blue
  hues.  The points that lie in the $\dot{M}$ range from $\sim$0.8 to
  $\sim$1.7 M$_{\sun}$/s correspond to simulations using the
  11.2-M$_{\sun}$ progenitor, and the points in the range from $\sim$0.2 to
  $\sim$3.5 M$_{\sun}$/s correspond to 15 M$_{\sun}$. The three resolutions are represented by stars (1),
  upside-down triangles (2), and squares (3).  The fit for $L_{\nu_e \bar{\nu}_e}$ as
  a function of $\dot{M}$ from \citet{burrows93} is plotted (solid
  line).  While this analytic expression passes through the 15.0-1D
  points it over predicts the critical luminosities for the
  11.2-1D runs by $\sim$15\%.  These results show that time-dependent
  1D and 2D simulations reproduce the critical $L_{\nu_e}$ and
  $\dot{M}$ condition, the critical luminosity is indeed a function of
  $\dot{M}$, and the critical luminosity for 2D
  simulations is $\sim$70\% the critical luminosity for 1D calculations.  \label{lumvsmdot}}
\end{figure}

Figure \ref{lumvsmdot} indicates that the concept of a
critical luminosity versus accretion rate condition is relevant for time-dependent
1D and 2D simulations and quantifies the difference in critical
luminosity between 1D and 2D simulations.
It plots electron- plus anti-electron-neutrino luminosity, $L_{\nu_e \bar{\nu}_e}$ ,
  vs. accretion rate, $\dot{M}$.
The trajectory of each model evolves from right to left
as the accretion rate decreases with time and the luminosity is held
constant.  For models that explode, we plot the luminosities and accretion rates
at explosion. These points strongly suggest $L_{\nu_e
  \bar{\nu}_e}$-$\dot{M}$ curves that separate exploding from
non-exploding models.  Indeed, models that do not explode remain in
the low luminosity and high-accretion-rate realm that is delineated by the
critical curves.
With error bars, we show the range of
  accretion rates encompassed by $t_{\rm exp} \pm 50$ ms.  In general, 1D simulations are represented by
  orange, 2D simulations by green hues, and 2D-90$^{\circ}$ runs by blue
  hues.  The three resolutions are represented by stars (1),
  upside-down triangles (2), and squares (3).  The fit for $L_{\nu_e \bar{\nu}_e}$ as
  a function of $\dot{M}$ from \citet{burrows93} is plotted (solid
  line).  While their fit passes through the 15.0-1D
  points, it overpredicts the critical luminosities for the
  11.2-1D runs by $\sim$15\%.  The critical luminosity for all 2D
  simulations is $\sim$70\% of the critical luminosity for 1D
  calculations.

It is interesting that the points for the 15.0-1D sequence are
consistent with the results of \citet{burrows93}.  However, we caution
against interpreting this coincidence too literally.  While we do
employ very similar heating and cooling terms, it is not a priori clear how
the many differences between our more realistic time-dependent models
should compare to the assumptions made in \citet{burrows93} to obtain
steady-state solutions.

Furthermore, we note that the slope of our results differ
from the slope of the \citet{burrows93} fit and the trend
is discontinuous between the 11.2 and 15 M$_{\sun}$ results.
The fit from \citet{burrows93} assumed a specific mass for the
PNS, while our simulations have masses that naturally
arise from the initial Fe-core mass and the subsequent accretion
rates.  Just after the Si/O interface accretes onto the
PNS, the mass of the PNS, $M_{\rm PNS}$, is $\sim$1.3 M$_{\sun}$ baryonic for
the 11.2-M$_{\sun}$ progenitor and $\sim$1.45 M$_{\sun}$ baryonic for the
15-M$_{\sun}$ progenitor.  Afterwards, the progenitor-dependent
accretion rates continue to increase the mass until
explosion.  Of the models from the 15.0-2D3 suite that exploded, the PNS
masses at $t_{\rm exp}$ are 1.63, 1.59, and 1.52 M$_{\sun}$ for
$L_{\nu_e} = 1.8$, 1.9, and 2.0.  In contrast, the masses from the
11.2-2D3 suite are 1.4, 1.38, 1.35, 1.33, and 1.32 M$_{\sun}$ for
$L_{\nu_e} = 0.8$, 0.9, 1.0, 1.1, and 1.2, respectively.  Assuming
that the explosion mechanism can be described by the transition from
steady-state accretion to a neutrino-driven wind (see
\citet{burrows87} and \citet{burrows93}, and work in 
preparation), we expect the critical luminosity to have
a mass dependence of $L_{\nu_e} \propto M_{\rm PNS}^{4/5}$.
Interestingly, this scaling with PNS mass resolves the
slope discrepancy and the discontinuity between the 15 and
11.2-M$_{\sun}$ models in Fig. \ref{lumvsmdot}.

Although the concept of a critical $L_{\nu_e}$ and $\dot{M}$ condition was derived
using steady-state accretion models \citep{burrows93},
Fig. \ref{lumvsmdot} shows that the condition is appropriate for
time-dependent 1D and 2D models.
We find that the critical luminosity for explosion is a
function of $\dot{M}$, and that it roughly follows the
trend suggested by \citet{burrows93}.
The critical luminosity for explosions in 2D is $\sim$70\% of
the luminosity required for explosions in 1D simulations.
While there are differences
between the 90$^{\circ}$ and 180$^{\circ}$ simulations and between simulations
with different resolutions, they roughly show
the same reduction in critical luminosity when we go to 2D.

\section{Conclusions}
\label{section:conclusions}
Since \citet{burrows93} published the critical luminosity condition for explosions by the neutrino mechanism, the
relevance of this condition for 1D and 2D time-dependent
simulations has remained unresolved.  Furthermore, recent simulations
have hinted that 2D simulations teeter on the verge of explosion, when
1D simulations completely fail.  One wonders if the concept of a
critical condition in 2D simulations can quantitatively explain the
trend toward successful explosions by the neutrino mechanism.

With 95 simulations that parameterize $L_{\nu_e}$, $\dot{M}$, resolution,
and dimensionality, we investigate the criteria for neutrino-driven
explosions. The results of which indicate that:
\begin{itemize}
\item {\bf A critical luminosity and accretion rate condition is observed
  in time-dependent 1D simulations.}  \citet{burrows93} used the success and failure of obtaining
  steady-state accretion solutions to suggest a critical luminosity and
  mass accretion rate condition for explosions.  Others have found a critical neutrino
  luminosity condition for explosion in time-dependent 1D simulations
  \citep{ohnishi06,buras06a}.  We note that a critical luminosity is applicable
  for time-dependent 1D
  simulations
  and that it is dependent upon $\dot{M}$, as suggested by \citet{burrows93}.
\item {\bf Radial oscillations are characteristic of the transition
    between 1D accretion (non-exploding) and exploding simulations.}
  Our results add to a growing body of literature that obtain radial
  oscillations near the critical luminosity
  \citep{mayle85,ohnishi06,buras06a}.
\item {\bf A critical $L_{\nu_e}$  and $\dot{M}$ condition distinguishes
    accretion and exploding models of time-dependent 2D simulations.}
\item {\bf Unlike 1D simulations, radial oscillations are not prominent
  near the critical luminosity of 2D simulations.  Rather non-radial
  SASI oscillations characterize 2D simulations, even for
  non-exploding models.}
\item {\bf We suggest that $\tau_{\rm adv}/\tau_q$
is a useful diagnostic for explosion, but may not be a rigorous
condition for explosion.}  Specifically, we found that some 1D models
strongly satisfied this condition several times before explosion, and
2D models that exploded exhibited trends toward explosion well before
this condition was met.
\item {\bf Several integral quantities such as heating rate, heating
    efficiency, and mass in the gain region grow with time in 1D and 2D
    exploding models, while they decline in 1D and 2D non-exploding
    models.}
  \citet{scheck08} compared these quantities for 1D and 2D simulations
  with the same neutrino luminosity and found declining evolutions in
  the 1D non-exploding model and increasing trends in the 2D
  exploding model.  We find similar trends.
  From these results, one might suggest that the SASI leads to
  increases in gain region mass and net heating rate, which in turn
  leads to explosion.  However, 2D simulations that do not explode
  despite having pronounced SASI oscillations show declining
  curves. Furthermore, 1D exploding models also show increases in
  heating rate and gain region mass at
  explosion.  Therefore, we conclude that the rise in these quantities
  near explosion are not a unique result of the SASI.  Instead, the upward evolution is correlated with
  explosion, whether it be in 1D or 2D simulations.  This, combined
  with the fact that the integral measures change from declining to
  increasing evolution near explosion, suggests that the upward trends
  are a consequence of explosion rather than a cause of explosion.
\item {\bf We suggest an improved measure of the residence time of matter in the gain region,
  $\tau_r$, that better elucidates the conditions leading to explosions.} While $\tau_{\rm adv}$ is a convenient measure of the timescale
  that a parcel of mass is exposed to net heating,
  we find $\tau_r$ a more informative measure for diagnosing
  the conditions and criteria leading to and during explosion in 2D simulations.  For
  example, by following the trajectories of particles that
  start outside and within the gain region, we note that much of the
  mass that accretes through the shock receives very little heating as
  it quickly advects onto the inner core.  On the other hand, there is
  a significant amount of mass within the gain region that has long
  residence times and receives a great deal of neutrino heating.  It
  is this latter material that experiences longer and longer
  residence times and prolonged heating that likely reverses accretion
  into explosion.
\item {\bf Resolution does affect the SASI and the post-shock flow in
    2D simulations, but hardly
  alters the critical luminosity and accretion rate condition.}  It
  noticeably alters the explosion time, but the correspondence between
  $t_{\rm exp}$ and resolution is not monotonic, nor is it easy to
  disentangle the trends.  However, the 2D critical luminosity
  vs. $\dot{M}$ curve is little affected by resolution.
\item {\bf The 90$^{\circ}$ simulations explode earlier than
  the 180$^{\circ}$ counterpart, but again the critical luminosity
  condition is hardly affected.}  We contrast this
with the results of \citet{buras06b} who found an explosion for
a 180$^{\circ}$ simulation, while a 90$^{\circ}$ simulation did not explode.
However, their 90$^{\circ}$ simulation is $\pm45^o$ of the
equator and suppresses $\ell= 1$ and $\ell = 2$ nonradial flow, the
dominant components of the SASI.  We simulated the region from the pole to the
equator, which suppresses $\ell = 1$ but allows $\ell = 2$.
In agreement with our results, \citet{marek07b} found earlier explosions for
90$^{\circ}$ compared to 180$^{\circ}$ simulations using the same
hydrodynamics and neutrino transport as \citet{buras06b} but a
90$^{\circ}$ grid similar to ours.
In any case, \citet{buras06b} speculated that
their 90$^{\circ}$ simulation was close to explosion and not very
different than the 180$^{\circ}$ simulation, and we show that the critical luminosity for
simulations using 90$^{\circ}$ and 180$^{\circ}$ are very similar.
\item {\bf The critical luminosity for explosions in 2D simulations is
    $\sim$70\% of
  the luminosity required for explosions in 1D simulations.}
  Irrespective of resolution or the angular size of the simulation, this
  conclusion holds.
\end{itemize}

By employing local heating and cooling algorithms, we avoid expensive
neutrino transport, but we also sacrifice accuracy.
In our simulations, heating and cooling by neutrinos are
decoupled from one another, which is not consistent with neutrino transport
calculations.
In addition, simulations using MGFLD \citep{dessart06} and
``ray-by-ray'' Boltzmann transport \citep{marek07} have neutrino
luminosities that decrease by $\sim$50\% from 100 ms to 300 ms
past bounce.
Our constant $L_{\nu_e}$ certainly does not reflect
this evolution in neutrino luminosities.
Furthermore, a large fraction of $L_{\nu_e}$ is derived from accretion at
  early times \citep{scheck06}.  In our simulations, $L_{\nu_e}$ is completely
  decoupled from such effects, and, hence, we ignore any feedback
  due to this coupling.  While our assumptions compromise the
  accuracy of the results, they enable the parameter study that we
  present, and we suspect that the general conclusions and trends regarding
  the critical luminosity condition are still
  relevant for simulations including detailed transport.

\citet{marek07} suggest that using a softer EOS in conjunction
with general relativistic gravity leads to more compact PNSs and more
favorable conditions for explosion; we include neither effect.
The soft EOS that accompanies explosion has an incompressibility of
nuclear matter, $K$, of 180 MeV, while the stiffer EOS, which does not
explode, has $K = 263$ MeV.  While this suggests an
interesting dependence of explosion on the EOS, we note that
laboratory experiments indicate that $K = 240 \pm 20$ MeV
\citep{shlomo06,lattimer07}, a value that is more consistent with the
non-explosive model.
Regardless, with a compact PNS, the post-shock flow is positioned deeper in the
potential, making it harder to explode.  All else being equal, this
should increase the critical luminosity.  However, \citet{marek07}
argue that the more compact PNS is hotter, emitting more energetic
neutrinos (though they are gravitationally redshifted) and at a higher
luminosity, which might compensate for the higher critical luminosity.  
Because we parameterize the luminosity and temperature in this study,
these effects are not addressed here and must await a more consistent
treatment.

Of the conclusions listed above, the most striking is that
time-dependent 1D and 2D simulations show a critical neutrino luminosity and
accretion rate condition for successful explosions.
Equally significant is
that the critical luminosity for 2D simulations is $\sim$70\% of the
critical luminosity for 1D
simulations.  This quantitatively supports
suspicions that multi-dimensional effects increase the likelihood of
successful explosions by the neutrino mechanism.  While none have explicitly explored these results,
the simulations of the past decade are consistent with this
conclusion, while also being intuitive and concise.  In this
sense, these results are expected and reassuring.  Remarkably, this
global condition for successful explosions, which was
informed by 1D steady-state solutions, survives despite the
complexities of time-dependent multi-dimensional simulations.

For all permutations of the core-collapse mechanism considered in this paper,
oscillations are ubiquitous
during the transition from accretion to explosion.
Interestingly, the character of these oscillations for 1D and 2D
simulations is quite different, and leads to distinctly different critical
luminosities.  Specifically, we suggest that the extra degree of freedom in 2D
simulations offers an easier route to
explosion via the SASI.  This suggests that for other than the
lightest progenitors, the core-collapse mechanism is inherently
multi-dimensional, and an accurate theory for the core-collapse
mechanism must be developed in the context of the multi-dimensional Universe.

\begin{figure}
\epsscale{0.68}
\plotone{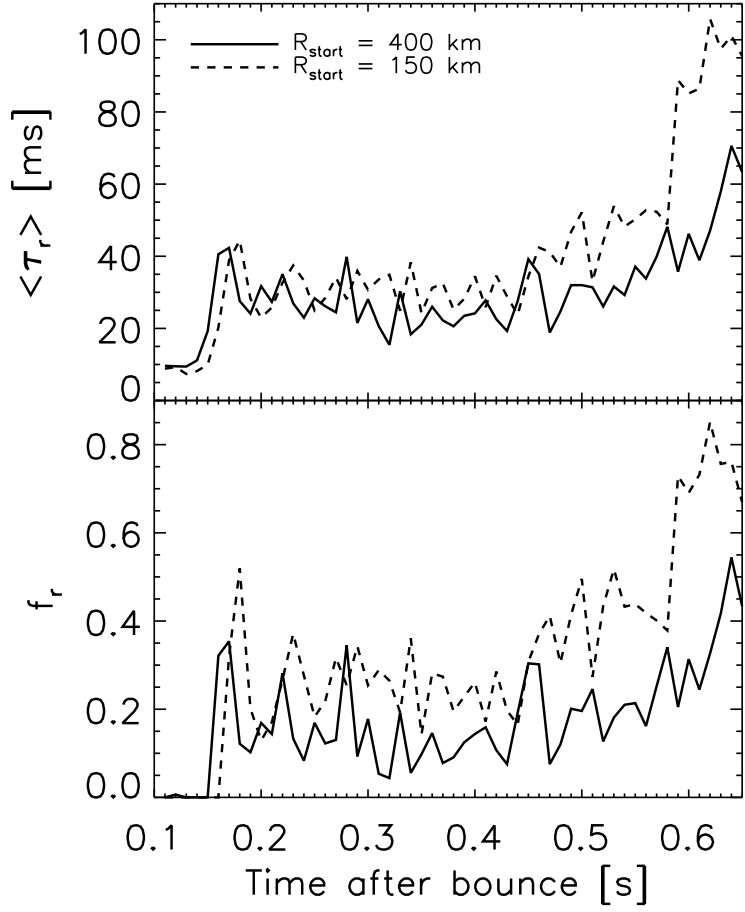}
\caption{The mean residence time ($\langle \tau_r \rangle$) (top
panel) and the fraction ($f_r$) of particles with $\tau_r > 40$ ms (bottom
panel) as a function of time for $R_{\rm start} = 400$ km (solid line)
and $R_{\rm start} = 150$ km (dashed line).  These residence times
were calculated for the 15.0-2D3, $L_{\nu_e} = 1.9$ model.  50,000 tracer particles are
initiated randomly in solid angle at either $R_{\rm start} = 150$ km
or $R_{\rm start} = 400$ km.  The trajectories for these particles are
integrated for 150 ms, and a new generation is initiated every 10 ms.
The time on the horizontal axis corresponds to the start time for a
generation.  From $\sim$200 ms and
beyond, particles that start in the gain region ($R_{\rm start} = 150$
km) consistently show larger $\langle \tau_r \rangle$ and more
particles with longer residence times.  There are four phases in this
plot that correspond to very short residence times, steady growth in
the SASI, an increase in the vigor of the SASI, and the initial
stages of explosion. See \S \ref{section:timescales} for a discussion
and Figs. \ref{stills1}, \ref{stills2}, and \ref{stills3} for entropy maps.
\label{taures_mean_frac}}
\end{figure}

\begin{figure}
\epsscale{0.6}
\plotone{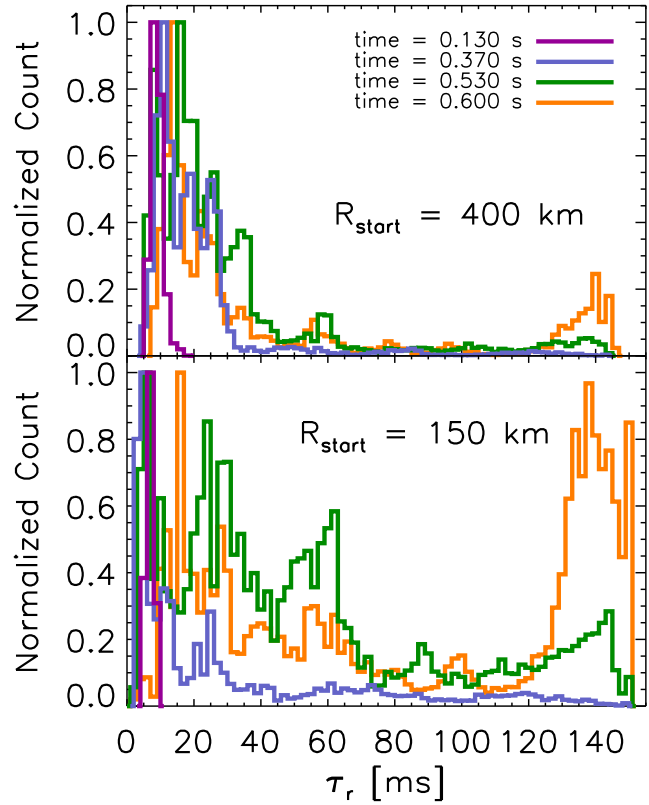}
\caption{The distribution of residence times,
$\tau_r$, for four generations of tracer particles corresponding to
the four phases in Fig. \ref{taures_mean_frac}.
  The times
shown in the legend correspond to the starting time for each
generation, and $R_{\rm start}$ indicates the starting radius for all
50,000 particles of each generation.  $R_{\rm start} = 400$ km is
situated well outside the shock radius, and $R_{\rm start} = 150$ km is approximately the middle
of the gain region.  The top panel shows that most of the mass that
accretes through the shock traverses the gain region on relatively
short timescales, $\sim$40 ms, while the bottom panel shows that
in addition to rapidly accreting plumes, there are regions in
the gain region in which mass has a large residence time and is
subjected to prolonged heating.  
\label{taures}}
\end{figure}

\begin{figure}
\epsscale{0.6}
\plotone{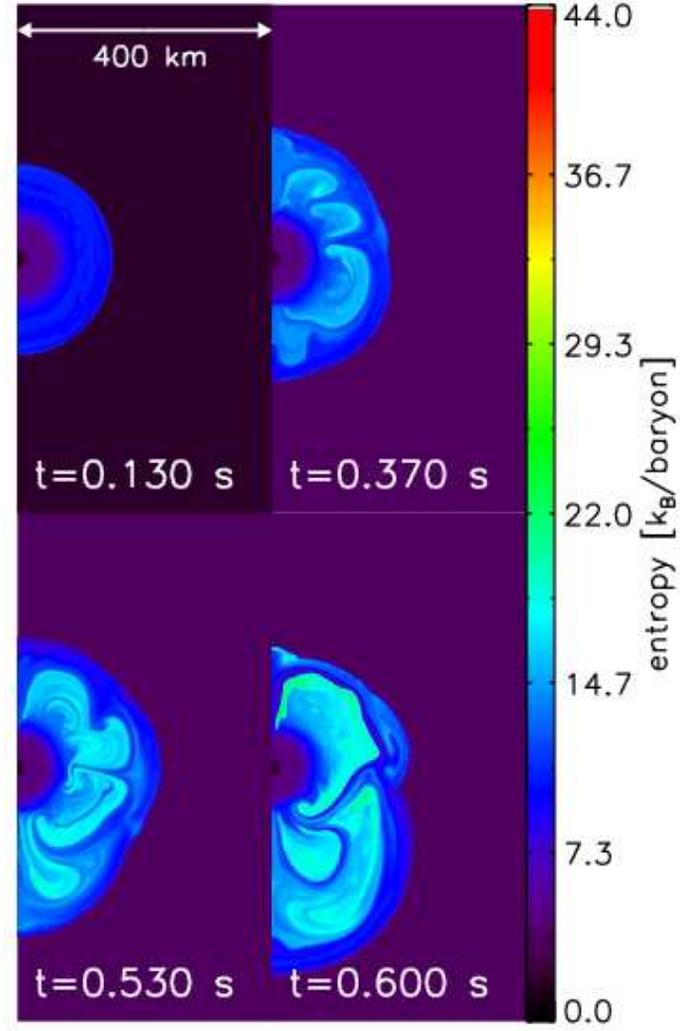}
\caption{Colormaps of entropy ($k_{\rm B}$/baryon) for 15.0-2D3,
  $L_{\nu_e} = 1.9$ and for the four phases
  shown in Fig. \ref{taures}.  Each panel is 400 km $\times$ 800 km.
  These snapshots of the flow correspond to the times in
  Fig. \ref{taures}, which label each histogram by the start time of
  the four generations of particles shown.  The four phases are:  1)
  there is very little convection or SASI and the particles accrete quickly ($\sim$9 ms) onto the
  PNS (top left), 2) the SASI undergoes a steady growth and
  the particles initiated at 400 km (200 km) have a mean residence time
  of 22 ms (32 ms) (top right), 3) the SASI is becoming quite vigorous and the
  residence times for both sets of particles are steadily increasing
  (bottom left), and 4) The SASI is very vigorous and 150 ms later
  (see Fig. \ref{stills3})
  the model undergoes explosion (bottom right). \label{stills1}}
\end{figure}

\begin{figure}
\epsscale{0.6}
\plotone{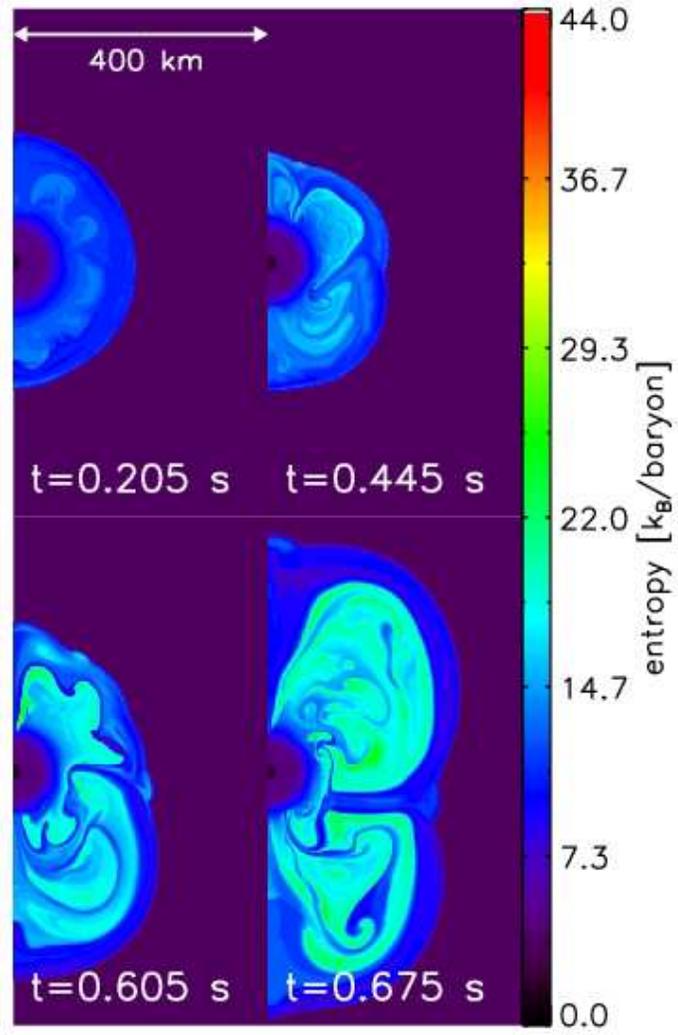}
\caption{The same as Fig. \ref{stills1}, except the flow is shown
  75 ms after the start of each generation of particles, which is the
  midpoint of particle tracking.  Low entropy plumes that quickly
  transport newly-accreted material to the PNS are quite pronounced in
  the bottom-right panel. \label{stills2}}
\end{figure}

\begin{figure}
\epsscale{0.6}
\plotone{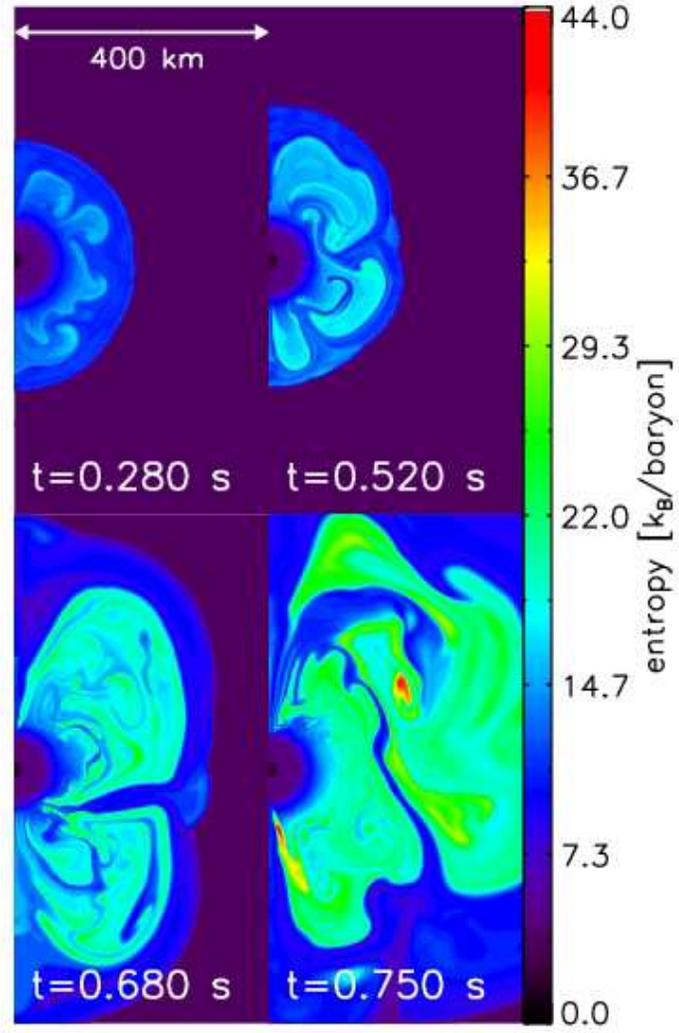}
\caption{The same as Fig. \ref{stills1}, except the flow is shown
  150 ms after the start of each generation of particles, which is the
  end of particle tracking. Note that the simulation is exploding in
  the bottom-right panel.\label{stills3}}
\end{figure}

\begin{figure}
\epsscale{0.775}
\plotone{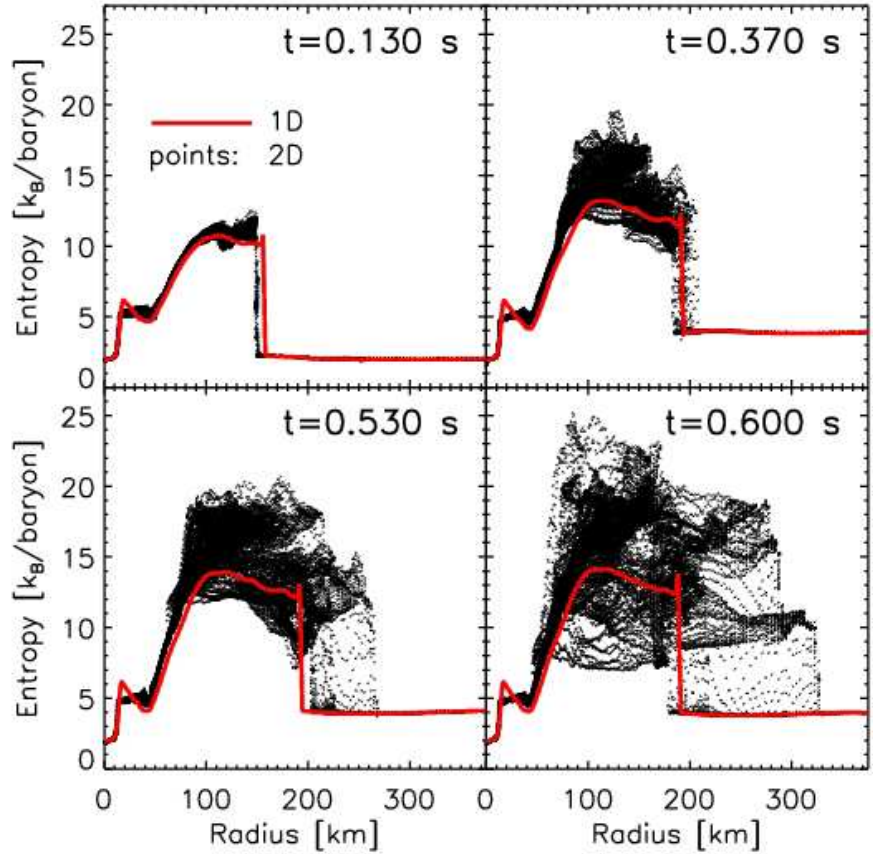}
\caption{Entropy vs. radius for the times shown in
  Fig. \ref{stills1}.  Solid red lines are entropy profiles for the
  15.0-1D, $L_{\nu_e} = 1.9$ simulation, and points show the entropy
  and radius for each zone of the 15.0-2D3, $L_{\nu_e} =
  1.9$ model.  These entropy profiles are coeval with the residence
  time distributions in Fig. \ref{taures} and the
  entropy maps in Fig. \ref{stills1}.  Features of the entropy
  profiles highlight the shock ($\sim$150 km for $t
  = 0.130$ s and $\sim$200 km for all other times), the gain region,
  ($\sim$100 km to $\sim$ 200 km), and the cooling
  region ($\sim$50 km to $\sim$100 km).  1D profiles show a negative entropy
  gradient in the gain region that results in convection in 2D
  simulations.  When the SASI dominates the post-shock flow, $t=
  0.370$, 0.530, and 0.600 s, the shock is distinctly aspherical, and a
  large range of entropies characterize the gain region.  Low
  entropies correspond to plumes that funnel matter onto the PNS,
  while high entropies correspond to regions of
  long residence times and more integrated net heating.  This is the
  likely cause for 2D simulations having lower critical luminosities
  than 1D simulations.
  \label{entropy}}
\end{figure}

Our results show that the critical luminosity for 2D
simulations is lower than the critical luminosity in 1D simulations,
but why is this the case?  The most obvious cause is
that the residence time of matter in the gain region is different for
1D and 2D simulations, and the longer residence times in 2D
simulations reduce the neutrino luminosity required for explosion.
So, what accounts for the different residence times when the accretion
rate is the same?  After all, we have shown that during the steady-state
accretion phase, the advection timescales determined by the flow
across the gain region ($\tau_{\rm adv}$, as calculated
by eq. (\ref{eq:tauadv})) and by $\tau_m = M_{\rm gain}/ \dot{M}$ are
similar.  Put another way, the solid-angle, average, radial velocity at radius $r$ is
\begin{equation}
v_r(r) = \frac{\dot{M}(r)}{4 \pi r^2 \rho(r)} \, ,
\end{equation}
and if $\rho(r)$ and $\dot{M}(r)$ are the same for 1D and 2D simulations,
the flow across the gain region should be similar.  Yet, 2D
simulations explode when 1D simulations do not.  This discrepancy can
be resolved by noting that the flow in the gain region is by no means
spherically symmetric in 2D simulations.  The plots, histograms,
and entropy maps of Figs. \ref{taures_mean_frac}-\ref{stills3} show
that much of the material that accretes through the shock is channeled
to plumes and quickly accrete onto the PNS.  At the same time, there
are large regions of high entropy material that have long residence
times in the gain region.  When the residence times within
these high entropy regions reach some threshold, it seems as though sufficient heating
has occurred to launch an explosion, and because relevant regions behind the
shock are in sonic contact, the shock radius expands globally.  Hence, the
critical luminosities for 2D simulations are lower because some matter
resides for longer times in the gain region for the same $\dot{M}$.

Since the concept of a critical luminosity condition applies for 1D and 2D
simulations, we assume that it should also apply for 3D
simulations.
Increasing the dimensionality from 1D to 2D, has proven favorable for successful explosions by the
neutrino mechanism.  Specifically, the
extra degree of freedom in 2D simulations increases the dwell time
of matter in the gain region.  Similarly, we suspect that differences between 2D
and 3D simulations will continue the trend toward lower critical
luminosities for higher dimensions.
For example, convection in 2D simulations experiences a turbulent energy cascade from
smaller to larger scales.
In Nature and in 3D simulations, however, the turbulent energy cascade is described
well by Kolmogorov theory, in which energy cascades from larger to
smaller scales.  If we analyze the transport of matter through the
convective gain region as a diffusive problem, then
the time to diffuse through the gain region would roughly scale as
\begin{equation}
\label{eq:tdiff}
t_{\rm diff} \sim d \left ( \frac{l^2}{\overline{v^2} \tau_{\rm coll} } \right ) \, ,
\end{equation}
where $d$ is the number of degrees of freedom, $l$ is the characteristic size, $\overline{v^2}$ is the mean-squared
speed of the diffusing particle, and $\tau_{\rm coll}$ is the average time
between collisions.
The extra degree of freedom in 3D simulations alone
could increase the dwell time by $\sim$50\% compared to 2D
simulations.  In addition, we suspect that the smaller scales of 3D
simulations would produce shorter $\tau_{\rm coll}$ compared to 2D
simulations and, consequently, further increase the dwell times.
Whether these scaling arguments bear out in realistic 3D simulations, in which
advection and the SASI are key in the flow, will be
determined by future 3D simulations.  If they are relevant, then the
increased dwell time might result in a critical luminosity in 3D
simulations that is even lower than the critical luminosity in 2D
simulations.  In turn, this might lead to successful explosions in 3D
radiation-hydrodynamic simulations.

\acknowledgments
We would like to thank Luc Dessart and Christian Ott for fruitful
comments.
 We acknowledge support for this work                                           
from the Scientific Discovery through Advanced Computing                        
(SciDAC) program of the DOE, under grant numbers DE-FC02-01ER41184              
and DE-FC02-06ER41452, and from the NSF under grant number AST-0504947.         
J.W.M. thanks the Joint Institute for Nuclear Astrophysics (JINA) for           
support under NSF grant PHY0216783.  In addition, J.W.M. is supported by an NSF Astronomy and Astrophysics Postdoctoral Fellowship under award AST-0802315.
We thank Jeff Fookson and Neal Lauver of the Steward Computer Support
Group for their invaluable help with the local Beowulf cluster, Grendel.

\bibliographystyle{apj}

\begin{thebibliography}{54}
\expandafter\ifx\csname natexlab\endcsname\relax\def\natexlab#1{#1}\fi

\bibitem[{{Bethe} \& {Wilson}(1985)}]{bethe85}
{Bethe}, H.~A., \& {Wilson}, J.~R. 1985, \apj, 295, 14

\bibitem[{{Bionta} {et~al.}(1987){Bionta}, {Blewitt}, {Bratton}, {Caspere}, \&
  {Ciocio}}]{bionta87}
{Bionta}, R.~M., {Blewitt}, G., {Bratton}, C.~B., {Caspere}, D., \& {Ciocio},
  A. 1987, Physical Review Letters, 58, 1494

\bibitem[{{Blondin} \& {Mezzacappa}(2006)}]{blondin06}
{Blondin}, J.~M., \& {Mezzacappa}, A. 2006, \apj, 642, 401

\bibitem[{{Blondin} {et~al.}(2003){Blondin}, {Mezzacappa}, \&
  {DeMarino}}]{blondin03}
{Blondin}, J.~M., {Mezzacappa}, A., \& {DeMarino}, C. 2003, \apj, 584, 971

\bibitem[{{Bruenn}(1985)}]{bruenn85}
{Bruenn}, S.~W. 1985, \apjs, 58, 771

\bibitem[{{Bruenn}(1989)}]{bruenn89}
---. 1989, \apj, 340, 955

\bibitem[{{Buras} {et~al.}(2006{\natexlab{a}}){Buras}, {Janka}, {Rampp}, \&
  {Kifonidis}}]{buras06b}
{Buras}, R., {Janka}, H.-T., {Rampp}, M., \& {Kifonidis}, K.
  2006{\natexlab{a}}, \aap, 457, 281

\bibitem[{{Buras} {et~al.}(2003){Buras}, {Rampp}, {Janka}, \&
  {Kifonidis}}]{buras03}
{Buras}, R., {Rampp}, M., {Janka}, H.-T., \& {Kifonidis}, K. 2003, Physical
  Review Letters, 90, 241101

\bibitem[{{Buras} {et~al.}(2006{\natexlab{b}}){Buras}, {Rampp}, {Janka}, \&
  {Kifonidis}}]{buras06a}
---. 2006{\natexlab{b}}, \aap, 447, 1049

\bibitem[{{Burrows} (1987)}]{burrows87}
{Burrows}, A. 1987, \apjl, 318, L57-L61

\bibitem[{{Burrows} {et~al.}(2007{\natexlab{a}}){Burrows}, {Dessart}, \&
  {Livne}}]{burrows07d}
{Burrows}, A., {Dessart}, L., \& {Livne}, E. 2007{\natexlab{a}}, in American
  Institute of Physics Conference Series, Vol. 937, Supernova 1987A: 20 Years
  After: Supernovae and Gamma-Ray Bursters, ed. S.~{Immler} \& R.~{McCray},
  370--380

\bibitem[{{Burrows} {et~al.}(2007{\natexlab{b}}){Burrows}, {Dessart}, {Livne},
  {Ott}, \& {Murphy}}]{burrows07c}
{Burrows}, A., {Dessart}, L., {Livne}, E., {Ott}, C.~D., \& {Murphy}, J.
  2007{\natexlab{b}}, \apj, 664, 416

\bibitem[{{Burrows} {et~al.}(2007{\natexlab{c}}){Burrows}, {Dessart}, {Ott}, \&
  {Livne}}]{burrows07b}
{Burrows}, A., {Dessart}, L., {Ott}, C.~D., \& {Livne}, E. 2007{\natexlab{c}},
  \physrep, 442, 23

\bibitem[{{Burrows} \& {Goshy}(1993)}]{burrows93}
{Burrows}, A., \& {Goshy}, J. 1993, \apjl, 416, L75+

\bibitem[{{Burrows} {et~al.}(1995){Burrows}, {Hayes}, \& {Fryxell}}]{burrows95}
{Burrows}, A., {Hayes}, J., \& {Fryxell}, B.~A. 1995, \apj, 450, 830

\bibitem[{{Burrows} {et~al.}(2006){Burrows}, {Livne}, {Dessart}, {Ott}, \&
  {Murphy}}]{burrows06}
{Burrows}, A., {Livne}, E., {Dessart}, L., {Ott}, C.~D., \& {Murphy}, J. 2006,
  \apj, 640, 878

\bibitem[{{Burrows} {et~al.}(2007{\natexlab{d}}){Burrows}, {Livne}, {Dessart},
  {Ott}, \& {Murphy}}]{burrows07a}
---. 2007{\natexlab{d}}, \apj, 655, 416

\bibitem[{{Burrows} {et~al.}(2000){Burrows}, {Young}, {Pinto}, {Eastman}, \&
  {Thompson}}]{burrows00}
{Burrows}, A., {Young}, T., {Pinto}, P., {Eastman}, R., \& {Thompson}, T.~A.
  2000, \apj, 539, 865

\bibitem[{{Dessart} {et~al.}(2006){Dessart}, {Burrows}, {Livne}, \&
  {Ott}}]{dessart06}
{Dessart}, L., {Burrows}, A., {Livne}, E., \& {Ott}, C.~D. 2006, \apj, 645, 534

\bibitem[{{Foglizzo}(2002)}]{foglizzo02}
{Foglizzo}, T. 2002, \aap, 392, 353

\bibitem[{{Foglizzo} {et~al.}(2007){Foglizzo}, {Galletti}, {Scheck}, \&
  {Janka}}]{foglizzo07}
{Foglizzo}, T., {Galletti}, P., {Scheck}, L., \& {Janka}, H.-T. 2007, \apj,
  654, 1006

\bibitem[{{Foglizzo} \& {Tagger}(2000)}]{foglizzo00}
{Foglizzo}, T., \& {Tagger}, M. 2000, \aap, 363, 174

\bibitem[{{Herant} {et~al.}(1994){Herant}, {Benz}, {Hix}, {Fryer}, \&
  {Colgate}}]{herant94}
{Herant}, M., {Benz}, W., {Hix}, W.~R., {Fryer}, C.~L., \& {Colgate}, S.~A.
  1994, \apj, 435, 339

\bibitem[{{Hirata} {et~al.}(1987){Hirata}, {Kajita}, {Koshiba}, {Nakahata}, \&
  {Oyama}}]{hirata87}
{Hirata}, K., {Kajita}, T., {Koshiba}, M., {Nakahata}, M., \& {Oyama}, Y. 1987,
  Physical Review Letters, 58, 1490

\bibitem[{{Janka}(2001)}]{janka01}
{Janka}, H.-T. 2001, \aap, 368, 527

\bibitem[{{Janka} \& {M\"{u}ller}(1995)}]{janka95}
{Janka}, H.-T., \& {M\"{u}ller}, E. 1995, \apjl, 448, L109

\bibitem[{{Janka} \& {M\"{u}ller}(1996)}]{janka96}
---. 1996, \aap, 306, 167

\bibitem[{{Kitaura} {et~al.}(2006){Kitaura}, {Janka}, \&
  {Hillebrandt}}]{kitaura06}
{Kitaura}, F.~S., {Janka}, H.-T., \& {Hillebrandt}, W. 2006, \aap, 450, 345

\bibitem[{{Lattimer} \& {Prakash}(2007)}]{lattimer07}
{Lattimer}, J.~M., \& {Prakash}, M. 2007, \physrep, 442, 109-165

\bibitem[{{Liebend{\"o}rfer}(2005)}]{liebendorfer05a}
{Liebend{\"o}rfer}, M. 2005, \apj, 633, 1042

\bibitem[{{Liebend{\"o}rfer} {et~al.}(2001{\natexlab{a}}){Liebend{\"o}rfer},
  {Mezzacappa}, \& {Thielemann}}]{liebendorfer01b}
{Liebend{\"o}rfer}, M., {Mezzacappa}, A., \& {Thielemann}, F.-K.
  2001{\natexlab{a}}, \prd, 63, 104003

\bibitem[{{Liebend{\"o}rfer} {et~al.}(2001{\natexlab{b}}){Liebend{\"o}rfer},
  {Mezzacappa}, {Thielemann}, {Messer}, {Hix}, \& {Bruenn}}]{liebendorfer01a}
{Liebend{\"o}rfer}, M., {Mezzacappa}, A., {Thielemann}, F.-K., {Messer}, O.~E.,
  {Hix}, W.~R., \& {Bruenn}, S.~W. 2001{\natexlab{b}}, \prd, 63, 103004

\bibitem[{{Liebend{\"o}rfer} {et~al.}(2005){Liebend{\"o}rfer}, {Rampp},
  {Janka}, \& {Mezzacappa}}]{liebendorfer05b}
{Liebend{\"o}rfer}, M., {Rampp}, M., {Janka}, H.-T., \& {Mezzacappa}, A. 2005,
  \apj, 620, 840

\bibitem[{{Livne}(1993)}]{livne93}
{Livne}, E. 1993, \apj, 412, 634

\bibitem[{{Livne} {et~al.}(2004){Livne}, {Burrows}, {Walder}, {Lichtenstadt},
  \& {Thompson}}]{livne04}
{Livne}, E., {Burrows}, A., {Walder}, R., {Lichtenstadt}, I., \& {Thompson},
  T.~A. 2004, \apj, 609, 277

\bibitem[{{Marek}(2007)}]{marek07b}
{Marek}, A. 2007, PhD thesis, Max Planck Institute for Astrophysics

\bibitem[{{Marek} \& {Janka}(2007)}]{marek07}
{Marek}, A., \& {Janka}, H.~. 2007, ArXiv e-prints, 708

\bibitem[{{Mayle}(1985)}]{mayle85}
{Mayle}, R.~W. 1985, PhD thesis, Lawrence Livermore National Laboratory,
  University of California, Berkeley

\bibitem[{{Mazurek}(1982)}]{mazurek82}
{Mazurek}, T.~J. 1982, \apjl, 259, L13

\bibitem[{{Murphy} \& {Burrows}(2008)}]{murphy08}
{Murphy}, J.~W., \& {Burrows}, A. 2008, submitted to \apj

\bibitem[{{Ohnishi} {et~al.}(2006){Ohnishi}, {Kotake}, \& {Yamada}}]{ohnishi06}
{Ohnishi}, N., {Kotake}, K., \& {Yamada}, S. 2006, \apj, 641, 1018

\bibitem[{{Ott} {et~al.}(2008){Ott}, {Burrows}, {Dessart}, \& {Livne}}]{ott08}
{Ott}, C.~D., {Burrows}, A., {Dessart}, L., \& {Livne}, E. 2008, ArXiv
  e-prints, 804

\bibitem[{{Rampp} \& {Janka}(2002)}]{rampp02}
{Rampp}, M., \& {Janka}, H.-T. 2002, \aap, 396, 361

\bibitem[{{Scheck} {et~al.}(2008){Scheck}, {Janka}, {Foglizzo}, \&
  {Kifonidis}}]{scheck08}
{Scheck}, L., {Janka}, H.-T., {Foglizzo}, T., \& {Kifonidis}, K. 2008, \aap,
  477, 931

\bibitem[{{Scheck} {et~al.}(2006){Scheck}, {Kifonidis}, {Janka}, \&
  {M{\"u}ller}}]{scheck06}
{Scheck}, L., {Kifonidis}, K., {Janka}, H.-T., \& {M{\"u}ller}, E. 2006, \aap,
  457, 963

\bibitem[{{Shen} {et~al.}(1998){Shen}, {Toki}, {Oyamatsu}, \&
  {Sumiyoshi}}]{shen98}
{Shen}, H., {Toki}, H., {Oyamatsu}, K., \& {Sumiyoshi}, K. 1998, Nuclear
  Physics A, 637, 435

\bibitem[{{Shlomo} {et~al.}(2006){Shlomo}, {Kolomietz}, \& {Col\`{o}}}]{shlomo06}
{Shlomo}, S., {Kolomietz}, V.~M., \& {Col\`{o}}, G. 2006,
{\it Eur. Phys. J.}, A30, 2

\bibitem[{{Thompson}(2000)}]{thompson00}
{Thompson}, C. 2000, \apj, 534, 915

\bibitem[{{Thompson} {et~al.}(2003){Thompson}, {Burrows}, \&
  {Pinto}}]{thompson03}
{Thompson}, T.~A., {Burrows}, A., \& {Pinto}, P.~A. 2003, \apj, 592, 434

\bibitem[{{Thompson} {et~al.}(2005){Thompson}, {Quataert}, \&
  {Burrows}}]{thompson05}
{Thompson}, T.~A., {Quataert}, E., \& {Burrows}, A. 2005, \apj, 620, 861

\bibitem[{{Wilson}(1985)}]{wilson85}
{Wilson}, J.~R. 1985 (Boston: Jones \& Bartlett), 422

\bibitem[{{Woosley} {et~al.}(2002){Woosley}, {Heger}, \& {Weaver}}]{woosley02}
{Woosley}, S.~E., {Heger}, A., \& {Weaver}, T.~A. 2002, Reviews of Modern
  Physics, 74, 1015

\bibitem[{{Woosley} \& {Weaver}(1995)}]{woosley95}
{Woosley}, S.~E., \& {Weaver}, T.~A. 1995, \apjs, 101, 181

\bibitem[{{Yamasaki} \& {Yamada}(2005)}]{yamasaki05}
{Yamasaki}, T., \& {Yamada}, S. 2005, \apj, 623, 1000

\bibitem[{{Yamasaki} \& {Yamada}(2006)}]{yamasaki06}
---. 2006, \apj, 650, 291

\end{thebibliography}


\clearpage

\begin{deluxetable}{cccccc}
\tabletypesize{\scriptsize}
\tablecaption{
Model parameters.\tablenotemark{1}
\label{table:sequences}
}
\tablewidth{0pt}
\tablehead{
  \colhead{Sequence Name} &
  \colhead{Mass (M$_{\sun}$)\tablenotemark{2}} &
  \colhead{Dimension\tablenotemark{3}} & 
  \colhead{$N_{r}$\tablenotemark{4}} &
  \colhead{$R_{\rm max}$ (km)\tablenotemark{5}}&
  \colhead{$L_{\nu_e}$ ($10^{52}$ erg s$^{-1}$)\tablenotemark{6}}}
\startdata
15.0-1D  & 15.0 & 1D & 700 & 4000 & 1.6-2.9 \\
15.0-2D1 & 15.0 & 2D & 250 & 4000 & 1.5-2.0 \\
15.0-2D2 & 15.0 & 2D & 400 & 4000 & 1.5-2.0 \\
15.0-2D3 & 15.0 & 2D & 400 & 1000 & 1.5-2.0 \\
15.0-Q1 & 15.0 & 2D-90$^{\circ}$ & 250 & 4000 & 1.5-2.0 \\
15.0-Q2 & 15.0 & 2D-90$^{\circ}$ & 400 & 4000 & 1.5-2.0 \\
15.0-Q3 & 15.0 & 2D-90$^{\circ}$ & 400 & 1000 & 1.5-2.0 \\
11.2-1D  & 11.2 & 1D & 700 & 4000 & 1.0-1.8 \\
11.2-2D1 & 11.2 & 2D & 250 & 4000 & 0.7-1.2 \\
11.2-2D2 & 11.2 & 2D & 400 & 4000 & 0.7-1.2 \\
11.2-2D3 & 11.2 & 2D & 400 & 1000 & 0.7-1.2 \\
11.2-Q1 & 11.2 & 2D-90$^{\circ}$ & 250 & 4000 & 0.7-1.2 \\
11.2-Q2 & 11.2 & 2D-90$^{\circ}$ & 400 & 4000 & 0.7-1.2 \\
11.2-Q3 & 11.2 & 2D-90$^{\circ}$ & 400 & 1000 & 0.7-1.2 \\
\enddata
\tablenotetext{1}{This table summarizes the 95 simulations presented
  in this paper. These
simulations represent a four-dimensional parameterization that
investigates the dependence of the conditions for explosion on the
accretion rate (column 2), dimensionality (column 3), resolution
(columns 4 \& 5), and neutrino luminosity (column 6).}
\tablenotetext{2}{Progenitor model.}
\tablenotetext{3}{Dimensionality.}
\tablenotetext{4}{Number of radial or effective radial zones.}
\tablenotetext{5}{Radius of the outer boundary}
\tablenotetext{6}{The range of neutrino luminosities investigated.}
\end{deluxetable}

\clearpage

\LongTables
\begin{deluxetable}{cccccc}
\tabletypesize{\scriptsize}
\tablecaption{
Conditions at the time of explosion.
\label{table:explosions}
}
\tablewidth{0pt}
\tablehead{ 
  \colhead{$L_{\nu_e}$ ($10^{52}$ erg s$^{-1}$)\tablenotemark{1}} &
  \colhead{$t_{\rm exp}$ (ms)\tablenotemark{2}} &
  \colhead{$\dot{M}_{\rm exp}$ (M$_{\sun}$/s)\tablenotemark{3}} &
  \colhead{$\dot{Q}$ (B/s)\tablenotemark{4}} &
  \colhead{$\dot{Q}/L_{\nu_e \bar{\nu}_e}$\tablenotemark{5}} &
  \colhead{$M_{\rm gain}$ (M$_{\sun}$)\tablenotemark{6}}}
\startdata
\cutinhead{11.2-1D\tablenotemark{7}}
1.3 & 932 & $0.084^{+0.006}_{-0.003}$ & 0.63 & 0.0244 & 0.0123 \\
1.4 & 496 & $0.112^{+0.012}_{-0.010}$ & 1.50 & 0.0535 & 0.0121 \\
1.5 & 312 & $0.136^{+0.025}_{-0.011}$ & 2.00 & 0.0668 & 0.0153 \\
1.6 & 409 & $0.124^{+0.002}_{-0.003}$ & 4.46 & 0.1393 & 0.0228 \\
\cutinhead{15.0-1D}
2.6 & 718 & $0.227^{+0.017}_{-0.015}$ & 3.69 & 0.0710 & 0.0105 \\
2.7 & 459 & $0.263^{+0.027}_{-0.013}$ & 3.79 & 0.0701 & 0.0129 \\
2.8 & 335 & $0.312^{+0.001}_{-0.006}$ & 5.57 & 0.0995 & 0.0177 \\
2.9 & 216 & $0.341^{+0.027}_{-0.027}$ & 3.42 & 0.0589 & 0.0319 \\
\cutinhead{15.0-2D1}
1.9 & 581 & $0.247^{+0.001}_{-0.001}$ & 1.49 & 0.0392 & 0.0123 \\
2.0 & 325 & $0.310^{+0.002}_{-0.001}$ & 1.97 & 0.0494 & 0.0167 \\
\cutinhead{15.0-2D2}
1.9 & 359 & $0.313^{+0.001}_{-0.023}$ & 1.51 & 0.0399 & 0.0120 \\
2.0 & 364 & $0.313^{+0.001}_{-0.027}$ & 1.60 & 0.0401 & 0.0133 \\
\cutinhead{15.0-2D3}
1.8 & 778 & $0.210^{+0.014}_{-0.007}$ & 1.21 & 0.0337 & 0.0099 \\
1.9 & 649 & $0.249^{+0.001}_{-0.015}$ & 1.33 & 0.0351 & 0.0113 \\
2.0 & 395 & $0.299^{+0.014}_{-0.031}$ & 1.76 & 0.0440 & 0.0152 \\
\cutinhead{15.0-Q1}
1.7 & 837 & $0.202^{+0.007}_{-0.005}$ & 0.91 & 0.0268 & 0.0074 \\
1.8 & 419 & $0.283^{+0.029}_{-0.024}$ & 1.19 & 0.0330 & 0.0098 \\
1.9 & 418 & $0.284^{+0.028}_{-0.024}$ & 1.31 & 0.0344 & 0.0109 \\
2.0 & 361 & $0.313^{+0.001}_{-0.025}$ & 1.71 & 0.0428 & 0.0152 \\
\cutinhead{15.0-Q2}
1.7 & 748 & $0.217^{+0.018}_{-0.009}$ & 0.91 & 0.0267 & 0.0073 \\
1.8 & 517 & $0.249^{+0.011}_{-0.002}$ & 1.30 & 0.0362 & 0.0119 \\
1.9 & 471 & $0.258^{+0.023}_{-0.010}$ & 1.19 & 0.0314 & 0.0100 \\
2.0 & 299 & $0.310^{+0.008}_{-0.001}$ & 1.54 & 0.0385 & 0.0126 \\
\cutinhead{15.0-Q3}
1.7 & 695 & $0.235^{+0.014}_{-0.018}$ & 1.14 & 0.0335 & 0.0096 \\
1.8 & 602 & $0.249^{+0.001}_{-0.001}$ & 1.08 & 0.0301 & 0.0089 \\
1.9 & 420 & $0.282^{+0.029}_{-0.024}$ & 1.42 & 0.0373 & 0.0122 \\
2.0 & 383 & $0.307^{+0.004}_{-0.033}$ & 1.66 & 0.0415 & 0.0135 \\
\cutinhead{11.2-2D1}
0.7 & 808 & $0.091^{+0.000}_{-0.001}$ & 0.23 & 0.0165 & 0.0050 \\
0.8 & 585 & $0.098^{+0.005}_{-0.002}$ & 0.30 & 0.0190 & 0.0055 \\
0.9 & 471 & $0.118^{+0.006}_{-0.013}$ & 0.43 & 0.0242 & 0.0068 \\
1.0 & 335 & $0.129^{+0.020}_{-0.005}$ & 0.55 & 0.0276 & 0.0086 \\
1.1 & 288 & $0.147^{+0.016}_{-0.019}$ & 0.89 & 0.0404 & 0.0131 \\
1.2 & 288 & $0.147^{+0.016}_{-0.019}$ & 0.89 & 0.0370 & 0.0131 \\
\cutinhead{11.2-2D2}
0.7 & 828 & $0.092^{+0.001}_{-0.002}$ & 0.24 & 0.0170 & 0.0054 \\
0.8 & 595 & $0.097^{+0.005}_{-0.002}$ & 0.31 & 0.0193 & 0.0053 \\
0.9 & 529 & $0.104^{+0.012}_{-0.005}$ & 0.40 & 0.0220 & 0.0070 \\
1.0 & 339 & $0.128^{+0.019}_{-0.004}$ & 0.54 & 0.0268 & 0.0077 \\
1.1 & 235 & $0.163^{+0.001}_{-0.014}$ & 0.85 & 0.0386 & 0.0115 \\
1.2 & 235 & $0.163^{+0.001}_{-0.014}$ & 0.85 & 0.0354 & 0.0115 \\
\cutinhead{11.2-2D3}
0.8 & 791 & $0.091^{+0.000}_{-0.001}$ & 0.32 & 0.0200 & 0.0057 \\
0.9 & 553 & $0.101^{+0.009}_{-0.004}$ & 0.40 & 0.0225 & 0.0059 \\
1.0 & 372 & $0.125^{+0.008}_{-0.001}$ & 0.60 & 0.0300 & 0.0087 \\
1.1 & 257 & $0.162^{+0.001}_{-0.025}$ & 0.91 & 0.0413 & 0.0122 \\
1.2 & 257 & $0.162^{+0.001}_{-0.025}$ & 0.91 & 0.0378 & 0.0122 \\
\cutinhead{11.2-Q1}
0.7 & 777 & $0.091^{+0.001}_{-0.001}$ & 0.23 & 0.0165 & 0.0048 \\
0.8 & 500 & $0.111^{+0.012}_{-0.009}$ & 0.31 & 0.0191 & 0.0060 \\
0.9 & 784 & $0.091^{+0.001}_{-0.001}$ & 0.78 & 0.0434 & 0.0133 \\
1.0 & 294 & $0.144^{+0.019}_{-0.017}$ & 0.52 & 0.0260 & 0.0076 \\
1.1 & 177 & $0.166^{+0.034}_{-0.004}$ & 0.91 & 0.0414 & 0.0124 \\
1.2 & 177 & $0.166^{+0.034}_{-0.004}$ & 0.91 & 0.0379 & 0.0124 \\
\cutinhead{11.2-Q2}
0.7 & 631 & $0.096^{+0.003}_{-0.002}$ & 0.22 & 0.0158 & 0.0040 \\
0.8 & 525 & $0.105^{+0.013}_{-0.006}$ & 0.28 & 0.0174 & 0.0056 \\
0.9 & 413 & $0.124^{+0.001}_{-0.004}$ & 0.39 & 0.0218 & 0.0071 \\
1.0 & 293 & $0.145^{+0.018}_{-0.017}$ & 0.54 & 0.0269 & 0.0075 \\
1.1 & 270 & $0.158^{+0.004}_{-0.025}$ & 0.82 & 0.0371 & 0.0107 \\
1.2 & 270 & $0.158^{+0.004}_{-0.025}$ & 0.82 & 0.0340 & 0.0107 \\
\cutinhead{11.2-Q3}
0.7 & 759 & $0.091^{+0.001}_{-0.001}$ & 0.22 & 0.0160 & 0.0047 \\
0.8 & 583 & $0.098^{+0.005}_{-0.003}$ & 0.30 & 0.0186 & 0.0056 \\
0.9 & 417 & $0.124^{+0.001}_{-0.005}$ & 0.39 & 0.0218 & 0.0065 \\
1.0 & 293 & $0.145^{+0.018}_{-0.017}$ & 0.56 & 0.0282 & 0.0083 \\
1.1 & 256 & $0.162^{+0.001}_{-0.024}$ & 0.84 & 0.0383 & 0.0125 \\
1.2 & 256 & $0.162^{+0.001}_{-0.024}$ & 0.84 & 0.0351 & 0.0125 \\
\enddata
\tablenotetext{1}{Electron-neutrino luminosity}
\tablenotetext{2}{The time of explosion}
\tablenotetext{3}{The mass accretion rate at explosion,}
\tablenotetext{4}{The net heating rate}
\tablenotetext{5}{The heating efficiency, where $L_{\nu_e \bar{\nu_e}}$ is the combined electron- and anti-electron-neutrino luminosity.}
\tablenotetext{6}{The mass in the gain region.}
\tablenotetext{7}{The table is divided into sections corresponding to the conditions at
explosion for each sequence presented in Table \ref{table:sequences}.}
\end{deluxetable}


\clearpage

\clearpage

\clearpage

\clearpage

\clearpage

\clearpage

\clearpage

\clearpage

\clearpage

\clearpage

\clearpage

\clearpage

\clearpage

\clearpage

\clearpage

\clearpage

\clearpage

\end{document}